\documentclass[fleqn,usenatbib]{mnras}

\usepackage{newtxtext,newtxmath}

\usepackage[T1]{fontenc}
\usepackage{ae,aecompl}

\usepackage{graphicx}	
\usepackage{amsmath}	
\usepackage{booktabs}
\usepackage{longtable}
\usepackage{subfig}
\usepackage{bm}
\usepackage{threeparttable}
\usepackage{xcolor}

\title[]{Ram Pressure Candidates in UNIONS}

\author[I. D. Roberts et al.]{
Ian D. Roberts$^{1}$\thanks{E-mail: iroberts@strw.leidenuniv.nl},
Laura C. Parker$^{2}$,
Stephen Gwyn$^3$,
Michael J. Hudson$^{4,5,6}$,
\newauthor
Raymond Carlberg$^7$,
Alan McConnachie$^3$,
Jean-Charles Cuillandre$^8$,
\newauthor
Kenneth C. Chambers$^9$,
Pierre-Alain Duc$^{10}$,
Hisanori Furusawa$^{11}$,
Raphael Gavazzi$^{12}$,
\newauthor
Vanessa Hill$^{13}$,
Mark E. Huber$^9$,
Rodrigo Ibata$^{10}$,
Martin Kilbinger$^8$,
Simona Mei$^{14}$,
\newauthor
Yannick Mellier$^{15}$,
Satoshi Miyazaki$^{11}$,
Masamune Oguri$^{16,17,18}$,
Richard J. Wainscoat$^9$
\\
\\
$^1$Leiden Observatory, Leiden University, PO Box 9513, 2300 RA Leiden, The Netherlands\\
$^2$Department of Physics and Astronomy, McMaster University, Hamilton ON L8S 4M1, Canada\\
$^3$NRC Herzberg Astronomy and Astrophysics, 5071 West Saanich Road, Victoria, BC V9E 2E7, Canada\\
$^4$Department of Physics \& Astronomy, University of Waterloo, Waterloo, ON N2L 3G1 Canada\\
$^5$Waterloo Centre for Astrophysics, University of Waterloo, 200 University Ave W, Waterloo, ON N2L 3G1, Canada\\
$^6$Perimeter Institute for Theoretical Physics, 31 Caroline St. North, Waterloo, ON N2L 2Y5, Canada\\
$^7$Department of Astronomy and Astrophysics, University of Toronto, Toronto, ON M5S 3H4, Canada\\
$^8$AIM, CEA, CNRS, Universit{\'e} Paris-Saclay, Universit{\'e} de Paris, F-91191 Gif-sur-Yvette, France\\
$^9$Institute for Astronomy, University of Hawaii, 2680 Woodlawn Drive, Honolulu, HI 96822, USA\\
$^{10}$Universit{\'e} de Strasbourg, CNRS, Observatoire astronomique de Strasbourg, UMR 7550, F‑67000 Strasbourg, France\\
$^{11}$National Astronomical Observatory of Japan, Mitaka, Tokyo 181-8588, Japan\\
$^{12}$Sorbonne Universit{\'e}, CNRS, UMR 7095, Institut d'Astrophysique de Paris, 98bis Bd Arago, F-75014 Paris\\
$^{13}$Universit{\'e} C{\^o}te d'Azur, OCA, CNRS, Lagrange, France\\
$^{14}$Universit{\'e} de Paris, CNRS, Astroparticule et Cosmologie, F-75013 Paris, France\\
$^{15}$Institut d’Astrophysique de Paris, UMR 7095 CNRS, Universit{\'e} Pierre et Marie Curie, 98bis boulevard Arago, F-75014, Paris, France\\
$^{16}$Kavli Institute for the Physics and Mathematics of the Universe (WPI), UTIAS, The University of Tokyo, Kashiwa, Chiba 277-8583, Japan\\
$^{17}$Department of Physics, The University of Tokyo, 7-3-1 Hongo, Bunkyo-ku, Tokyo 113-0033, Japan\\
$^{18}$Research Center for the Early Universe, The University of Tokyo, 7-3-1 Hongo, Bunkyo-ku, Tokyo 113-0033, Japan
}

\date{Accepted XXX. Received YYY; in original form ZZZ}

\pubyear{2020}

\begin{document}
\label{firstpage}
\pagerange{\pageref{firstpage}--\pageref{lastpage}}
\maketitle

\begin{abstract}
We present a search for disturbed, candidate ram pressure stripping galaxies across more than 50 spectroscopically selected SDSS groups and clusters.  Forty-eight ram pressure candidates are visually identified in these systems using high quality UNIONS imaging from the Canada-France Hawaii Telescope, covering $\sim\!6200\,\mathrm{deg^2}$ and $\sim\!2800\,\mathrm{deg^2}$ in the $u$- and $r$-bands respectively. Ram pressure candidates are found in groups and clusters spanning a wide range in halo mass and include $\sim\!30$ ram pressure candidates in the group regime ($M_h < 10^{14}$).  The observed frequency of ram pressure candidates shows substantial scatter with group/cluster mass, but on average is larger in clusters ($M_h \ge 10^{14}\,\mathrm{M_\odot}$) than groups ($M_h < 10^{14}\,\mathrm{M_\odot}$) by a factor of $\sim$2.  We find that ram pressure candidates are most commonly low-mass galaxies and have enhanced star formation rates relative to star-forming field galaxies.  The enhancement in star formation is largely independent of galaxy mass and strongest for galaxies in clusters.  As a result of the large survey footprint and excellent image quality from UNIONS, we are able to identify disturbed galaxies, potentially affected by ram pressure stripping, across a wide range of host environment.
\end{abstract}




\section{Introduction} \label{sec:intro}

Galaxy clusters host populations of galaxies with properties that clearly differ from galaxies in low-density environments.  Relative to galaxies in low-mass groups, or isolated in the field, galaxy populations in clusters are systematically redder, more gas poor, and have enhanced quiescent fractions.  This prevalence of `red and dead' galaxies in dense environments was noted by early, formative papers \citep[e.g.][]{dressler1980} and has since been confirmed by large, modern galaxy surveys \citep[e.g.][]{wetzel2012,haines2015,brown2017}.  As these observed environmental trends have become more clear, the focus has shifted toward a physical understanding of the mechanisms causing these galaxy transformations.  In particular, substantial effort has been devoted to understanding which physical processes are responsible for shutting off star formation (`quenching') in cluster galaxies.
\par
The cluster environment subjects galaxies to gravitational/dynamical interactions with other member galaxies, as well as hydrodynamic interactions with the hot ($T \sim 1-10\,\mathrm{keV}$, e.g. \citealt{mohr1999}) intracluster medium.  These interactions are unique to dense environments and likely drive the strong environmental dependence of galaxy properties which is observed.  Specific physical processes capable of quenching star formation in clusters have been proposed, all of which involve gravitational and/or hydrodynamic interactions.  Examples of these quenching mechanisms include: directly stripping gas from galaxy discs (ram-pressure stripping, tidal stripping, viscous stripping, e.g. \citealt{gunn1972,nulsen1982,quilis2000}), preventing gas cooling within galaxies (starvation/strangulation, e.g. \citealt{larson1980,peng2015}), or inducing strong starbusts, typically as gas is funneled toward galaxy centres, which quickly consume cold-gas reserves (mergers, galaxy harassment, e.g. \citealt{mihos1994a,mihos1994b,moore1996}).
\par
The build-up of massive clusters is a hierarchical process involving mergers with other galaxy clusters as well as lower mass groups. Therefore, some passive cluster galaxies may have actually been quenched in a lower mass group prior to accretion onto the more massive host cluster.  This scenario is referred to as pre-processing and numerous works have attempted to constrain the relative contribution of pre-processing to the cluster red sequence \citep[e.g.][]{mcgee2009,vonderlinden2010,haines2015,roberts2017,pallero2020}.  It is also likely that the efficiency of different quenching mechanisms differs between massive clusters and lower mass groups.  For example, galaxy mergers are more common in groups where relative velocities are smaller, whereas ram pressure stripping should be stronger in massive clusters with dense ICMs and large velocity dispersions \citep[e.g.][]{hickson1997,hester2006,rasmussen2006,darg2010}.  Large redshift surveys at low-$z$, which contain thousands of groups and clusters across a wide range in mass ($\sim\!10^{13} - 10^{15}\,\mathrm{M_\odot}$), provide an excellent avenue for constraining this pre-processing.
\par
Of key importance is identifying signatures of specific quenching mechanisms which can be observed and used to disentangle the relative contributions of various processes.  For galaxies undergoing strong ram pressure stripping, they are expected to leave a wake of stripped gas opposite to the direction of motion with respect to the ICM.  Furthermore, bow shocks and enhanced star formation are expected on the galaxy leading edge \citep[e.g.][]{clemens2000,yun2019}.  Typically, these stripped features are observed with the 21 cm line from neutral atomic hydrogen \citep[e.g.][]{kenney2004,chung2007,chung2009,kenney2015} or the $\mathrm{H}\alpha$ line tracing (partially) ionized gas \citep[e.g.][]{poggianti2017,boselli2018}.  That said, stripped features have also been observed in molecular hydrogen \citep[e.g.][]{vollmer2012,lee2017,lee2018,jachym2019}, dust \citep[e.g.][]{crowl2005}, far-ultraviolet \citep[e.g.][]{smith2010,boissier2012,george2018}, X-rays \citep[e.g.][]{sun2010,poggianti2019_baryon,sun2021}, and the radio continuum \citep[e.g.][]{gavazzi1987,vollmer2009,chen2020,roberts2021_LOFARclust}.  While these methods provide an accurate means for identifying galaxies experiencing stripping, the required observations are often expensive and therefore target galaxies are typically known \textit{apriori} to be morphologically disturbed.  Alternatively, it is possible to identify galaxies which are likely undergoing ram pressure stripping with rest-frame optical imaging \citep{mcpartland2016,poggianti2016,roberts2020,durret2021}.  In particular, blue filters (such as $u$-band) can be used to identify disturbed morphological features associated with ram pressure stripping \citep{smith2010}.  These features are typically identified by-eye, but these visual classifications have been shown to agree well with quantitative morphological measures such as Gini-$\mathrm{M_{20}}$ and concentration- asymmetry \citep{conselice2003,lotz2004,mcpartland2016,roberts2020}.  Since broad-band imaging is primarily tracing stellar-light, and the galaxy stellar components are less perturbed by ram pressure stripping than neutral and ionized gas, ram pressure classifications from optical imaging will be less accurate and/or less complete than corresponding \textsc{Hi} or $\mathrm{H\alpha}$ surveys.  That said, the advantage of broad band imaging is the large areas which can be surveyed compared to spectroscopic observations, allowing for an unbiased, blind search for ram pressure stripped galaxies across many groups and clusters.
\par
In this work we take advantage of the extremely large survey footprint ($\sim$thousands of square degrees of the northern sky) and high resolution imaging (sub-arcsecond image quality) from the Ultraviolet Near-Infrared Optical Northern Survey (UNIONS).  This multiband imaging survey includes the deep $u$- and $r$-band photometry from the Canada France Imaging Survey (CFIS) that is used in this work.  Photometry in blue and/or near-UV filters is valuable for visually identifying candidate ram pressure galaxies \citep{roberts2020}, as it traces a combination of young stellar populations and ionized gas.  By matching the UNIONS footprint to SDSS spectroscopy, we perform an unbiased search for ram pressure stripping across more than $50$ groups and clusters with $\sim\!1000$ star-forming, spectroscopic members.  The size of this sample allows the investigation of the frequency of ram pressure stripping as a function of both galaxy stellar mass and group/cluster mass.
\par
The layout of this paper is the following.  In Section~\ref{sec:data} we describe the UNIONS imaging as well as the SDSS group/cluster and field galaxy samples used in this work.  In Section~\ref{sec:RPcandidates} we describe the procedure that we employ to identify candidate ram pressure galaxies.  In Section~\ref{sec:results} we present the primary results of this work including the frequency of ram pressure candidates as a function of galaxy stellar mass and group/cluster mass (\ref{sec:RP_demographics}), the position of ram pressure candidates in projected phase space and the orientation of observed asymmetries (\ref{sec:PS}), and the star formation properties of ram pressure candidates relative to `normal' star-forming cluster and field galaxies (\ref{sec:star_formation}).  We discuss these results in Section~\ref{sec:discussion} and give our main conclusions in Section~\ref{sec:conclusion}.
\par
Throughout the paper we assume a flat, $\mathrm{\Lambda}$ cold dark matter cosmology with $\Omega_\mathrm{M} = 0.3$, $\Omega_\mathrm{\Lambda} = 0.7$, and $H_0 = 70\,\mathrm{km\,s^{-1}\,Mpc^{-1}}$.

\section{Data} \label{sec:data}

\subsection{UNIONS Imaging} \label{sec:cfis_imaging}

UNIONS is a new consortium of wide field imaging surveys of the northern hemisphere. UNIONS consists of CFIS, members of the Pan-STARRS team, and the Wide Imaging with Subaru HyperSuprimeCam of the Euclid Sky (WISHES) team. Each team is currently collecting imaging at their respective telescopes: CFHT/CFIS is targeting deep $u$- and $r$- band photometry, Pan-STARRS is obtaining deep $i$- and moderate-deep $z$-bands, and Subaru/WISHES is acquiring deep $z$. These independent efforts are directed, in part, to securing optical imaging to complement the \textit{Euclid} space mission, although UNIONS is a separate consortium aimed at maximizing the science return of these large and deep ground-based surveys of the northern skies. In the current analysis, we use the UNIONS/CFIS $u$- and $r$-band data only.
\par
CFIS \citep{ibata2017} is a Canada-France Hawaii Telescope (CFHT) Large Program which will image the northern sky in $u$- and $r$-band over $\sim\!10\,000\,\mathrm{deg^2}$ and $5000\,\mathrm{deg^2}$ respectively.  $r$-band imaging will be obtained above a declination of $30^\circ$ and the $u$-band imaging will cover declinations above $0^\circ$.  At the time of writing, $\sim\!6200\,\mathrm{deg^2}$ of $u$-band imaging and $\sim\!2800\,\mathrm{deg^2}$ of $r$-band imaging has been collected and reduced.  CFIS provides sub-arcsecond image quality (IQ) and reaches depths typically 3 mag deeper than SDSS photometry. The $u$- and $r$-band imaging in this work is taken with the MegaCam wide-field imaging instrument at the CFHT, which covers a full $1 \times 1\,\mathrm{deg^2}$ field-of-view with a pixel scale of $0.187\,\mathrm{''/px}$.  Raw images are flat fielded and bias subtracted, astrometry is calibrated with Gaia DR2 \citep{gaia2016,gaiaDR2}, and photometric calibration is done with Pan-STARRs \citep{chambers2016}.  Individual images are stacked onto $0.5 \times 0.5\,\mathrm{deg}$ tiles, and images used in this work are extracted from these tiles.  The median IQ across all CFIS imaging is $0.92''$ and $0.68''$ for the $u$- and $r$-bands, and the median depths ($5\sigma$ point source) are 24.5 and 25.3 for $u$ and $r$.  This excellent image quality, which is relatively uniform across the sky, is an asset which makes it possible to identify ram pressure candidates (see Section~\ref{sec:RPcandidates}) across a wide range in galaxy mass and environment.

\subsection{Cross-matching UNIONS with SDSS Groups and Clusters} \label{sec:sdss_match}

\begin{figure}
    \centering
    \includegraphics[width=\columnwidth]{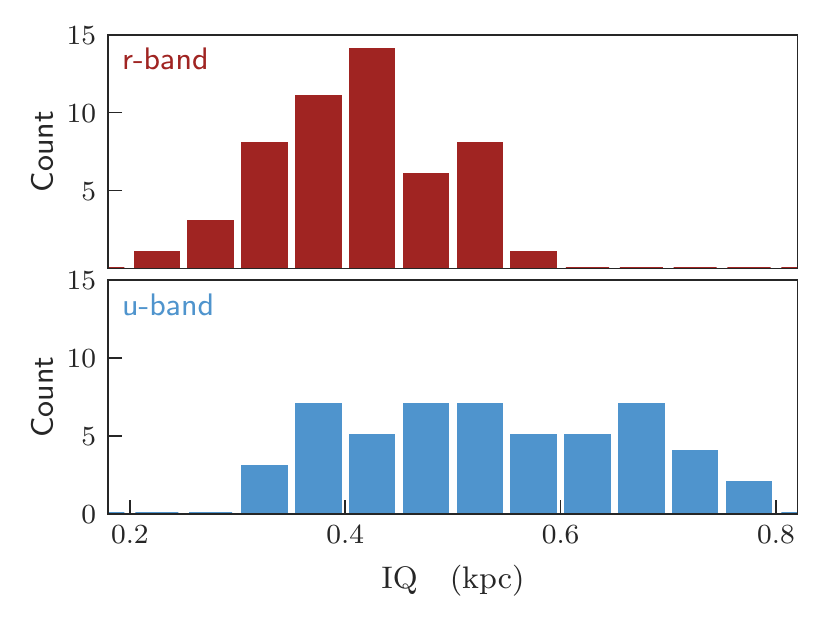}
    \\[0.5em]
    \includegraphics[width=\columnwidth]{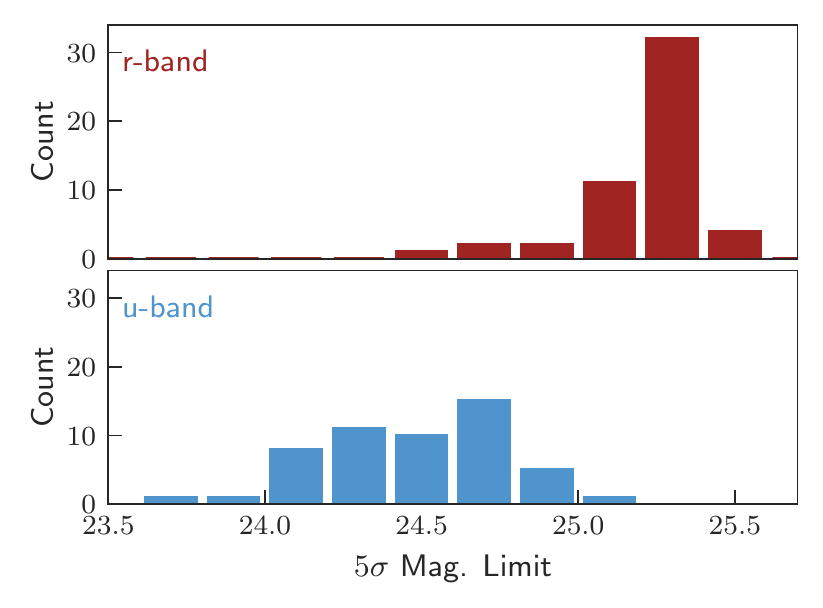}
    \caption{\textit{Top:}  Median $u$- and $r$- band image quality, in kpc, for each of the CFIS clusters in the sample.  \textit{Bottom:}  Median $5\sigma$ $u$- and $r$-band magnitude limits for each of the CFIS clusters in the sample.}
    \label{fig:IQ_depth}
\end{figure}

To complement the imaging provided by UNIONS, we cross-match the CFIS footprint to galaxies with SDSS spectroscopic redshifts, focusing on galaxies in the vicinity of SDSS groups and clusters.  We use the \citet{lim2017} SDSS group catalogue, which follows the group finding algorithm from the \citet{yang2005,yang2007}, with an improved method for halo mass assignment. We select all groups and clusters with 7 or more spectroscopic members, halo masses greater than $10^{13}\,h^{-1}\,\mathrm{M_\odot}$, and $u$- and $r$-band CFIS imaging.  Seven galaxies corresponds to the typical galaxy membership for groups in the \citeauthor{lim2017} catalog at $z<0.04$ with $M_h \sim 10^{13}\,\mathrm{M_\odot}$. This also still gives enough spectroscopic members to derive reasonably accurate luminosity-weighted centres.  We only include groups/clusters at $z<0.04$ to ensure sub-kpc image quality for our entire galaxy sample.  In Fig.~\ref{fig:IQ_depth} (top) we show the distributions of median image quality (IQ) for each group/cluster in both $u$- and $r$-band.  The IQ is best for the $r$-band observations, but for both $u$- and $r$- band the median IQ is sub-kpc for all groups/clusters.  In Fig.~\ref{fig:IQ_depth} (bottom) we show the median $5\sigma$ magnitude limits for each group/cluster.  The $u$-band limiting magnitudes range between 24th and 25th magnitude and the $r$-band limiting magnitudes are almost all fainter than 25th magnitude.
\par
To select member galaxies for each group/cluster we do not use the galaxy memberships given in \citeauthor{lim2017} catalogue,  and instead include all galaxies which are with $1 \times R_{180}$ and $3 \times \sigma_v$ of the group/cluster centre and redshift given in \citet{lim2017}.  This is a loose membership criteria, which likely includes some galaxies not strictly bound to their host system.  We opt for this approach to ensure that we do not exclude galaxies just starting their infall, and note that the qualitative results of this work are unchanged with the stricter galaxy memberships given by \citet{lim2017}.  For each group/cluster the virial radius, $R_{180}$, and velocity dispersion, $\sigma_v$, are calculated following \citet{lim2017} as
\begin{equation}
    R_{180} = 1.33\,h^{-1}\,\mathrm{Mpc}\,\left(\frac{M_h}{10^{14}\,h^{-1}\,\mathrm{M_\odot}}\right)^{1/3} (1+z_\mathrm{cluster})^{-1}
\end{equation}

\begin{equation}
    \sigma_v = 418\,\mathrm{km\,s^{-1}}\,\left(\frac{M_h}{10^{14}\,h^{-1}\,\mathrm{M_\odot}}\right)^{0.3367}
\end{equation}
\noindent
where $M_h$ is the halo mass and $z_\mathrm{cluster}$ is the cluster redshift.  The criteria outlined above select 69 groups and clusters with halo masses ranging from $10^{13.1}\,\mathrm{M_\odot}$ to $10^{14.4}\,\mathrm{M_\odot}$.  With our loose membership criteria, we identify a total of 2059 member galaxies with SDSS redshifts across these systems.  For all galaxies we obtain stellar masses ($M_\mathrm{star}$) and star formation rates (SFRs) from the GSWLC-2 catalogue \citep{salim2016,salim2018}, which derives stellar masses and SFRs for SDSS galaxies via SED fitting from the UV to the mid-IR.  To ensure uniform coverage for the galaxies in our sample, we use the shallow, all-sky catalog.  For the remainder of the paper, we define star-forming galaxies to be all galaxies with specific star formation rates above $10^{-11}\,\mathrm{yr^{-1}}$ ($\mathrm{sSFR} = \mathrm{SFR} / M_\mathrm{star}$).  Of the member galaxies, 989 are star-forming by this criteria.

\subsection{Field Sample} \label{sec:field_sample}

We use the SDSS isolated field galaxy sample from \citet{roberts2017}.  In brief, this field sample is compiled by selecting all galaxies in single-member groups from the \citet{yang2005,yang2007} catalogue, which are isolated from the nearest `bright' galaxy by at least $1\,\mathrm{Mpc}$ and $1000\,\mathrm{km\,s^{-1}}$.  Bright galaxies are defined to be any galaxy with an $r$-band absolute magnitude brighter than the SDSS magnitude limit at $z=0.04$ (the redshift upper limit of this sample).  We further restrict the field sample by only including galaxies which overlap with both $u$- and $r$- band imaging tiles from the CFIS survey.  This gives a field sample of $1352$ SDSS galaxies with multiband CFIS imaging.  Stellar masses and SFRs for the field sample are also obtained from the GSWLC-2 catalogue \citep{salim2016,salim2018}.  The stellar mass and redshift distributions for the field sample  are well matched to the sample of group galaxies described in the previous section.

\section{Identifying Ram Pressure Candidates} \label{sec:RPcandidates}

\begin{figure*}
    \centering
    \includegraphics[width=0.9\textwidth]{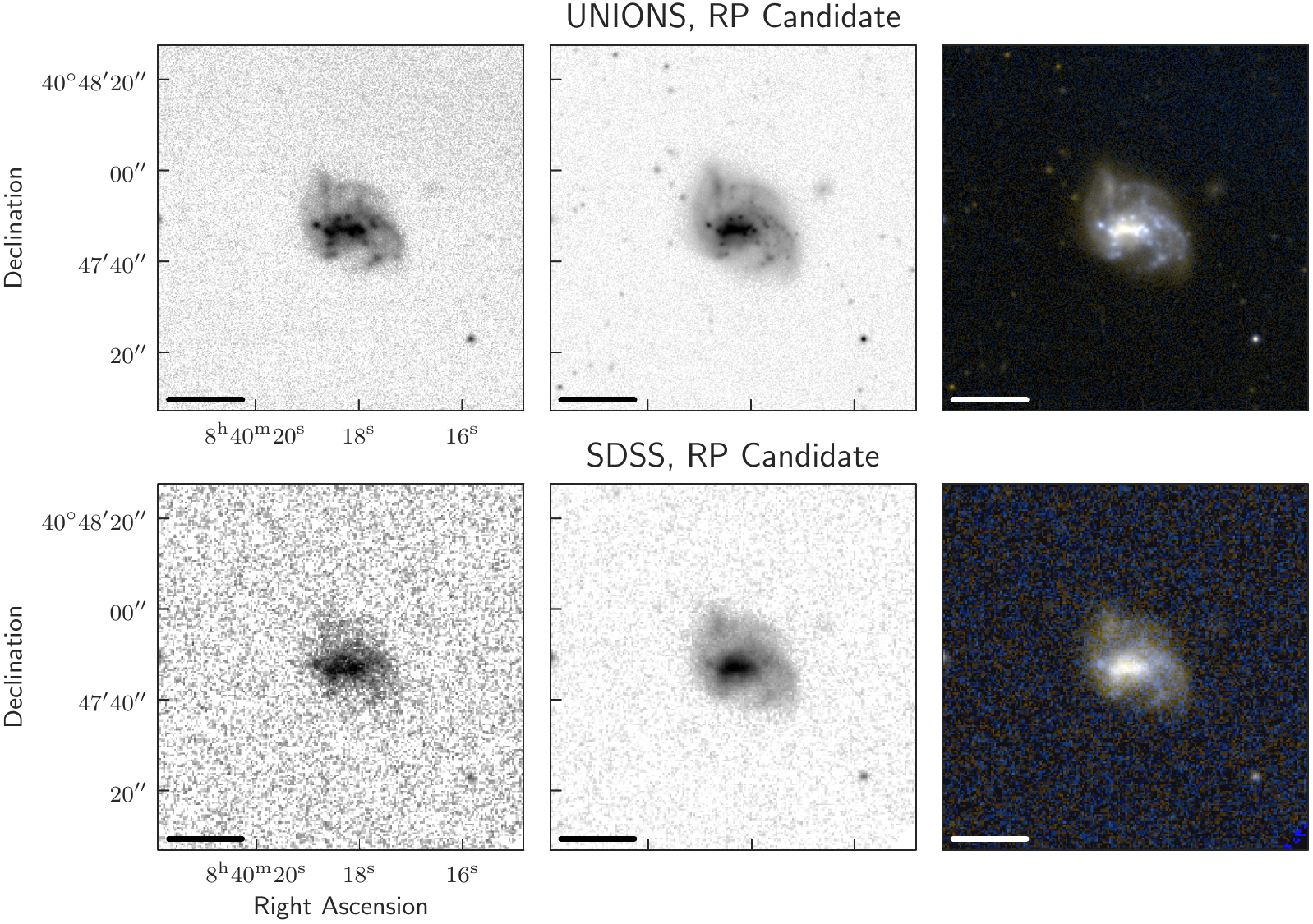}
    \includegraphics[width=0.9\textwidth]{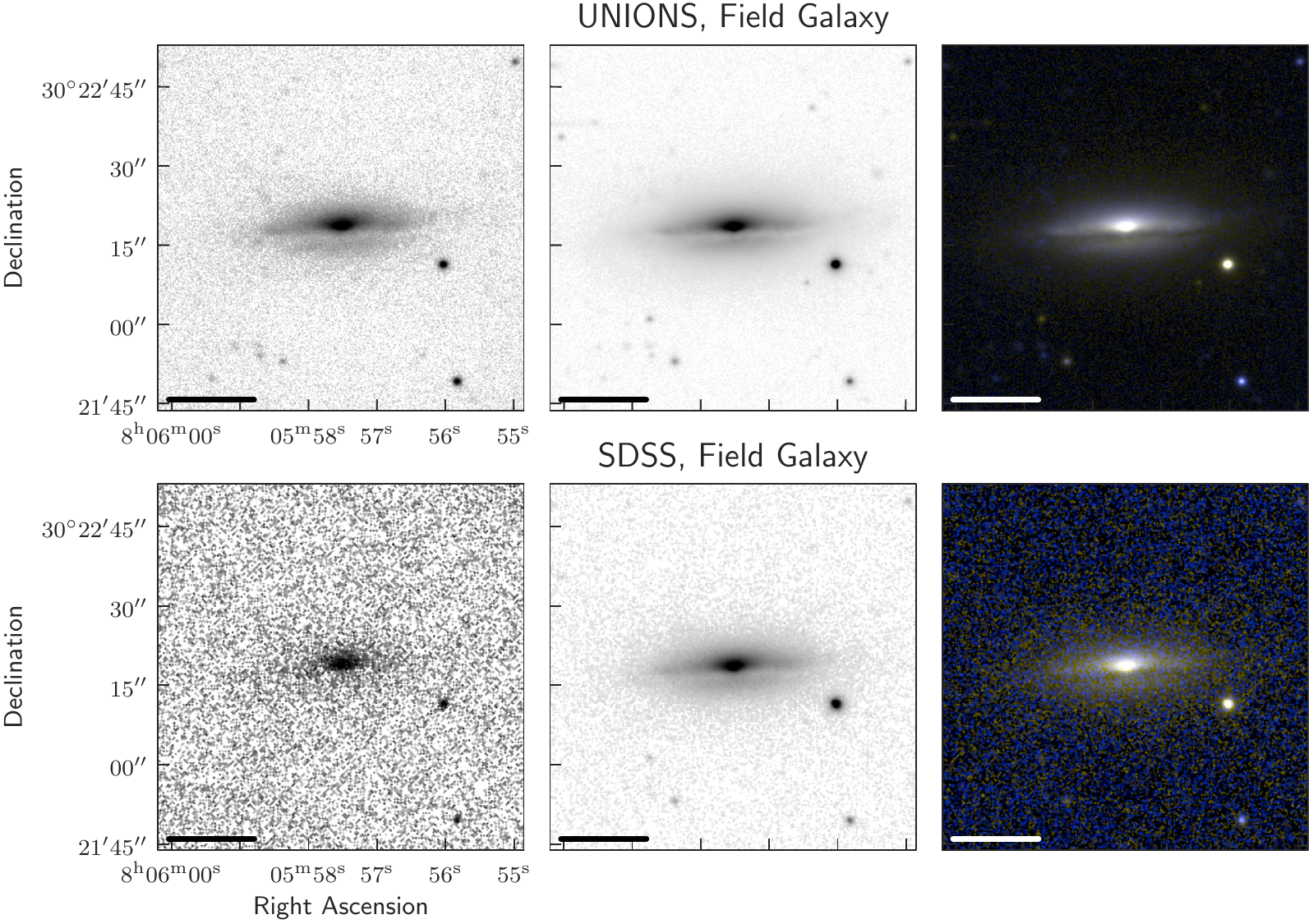}
    \caption{Example $u$-band, $r$-band, and rgb images of a ram pressure candidate and a randomly selected field galaxy.  For each, in the top row we show the UNIONS imaging used for classification in this work, and for comparison in the bottom row we show the analogous images from the SDSS, highlighting the excellent image quality in UNIONS.  The scale bar in each panel corresponds to a physical size of $10\,\mathrm{kpc}$.}
    \label{fig:RPexamples}
\end{figure*}

We identify ram pressure candidates from pseudo three-colour images.  For each star-forming galaxy an rgb thumbnail is made with the $r$-band in the r-channel, the $u$-band in the b-channel, and the average of the $u$- and $r$-bands in the g-channel.  $50 \times 50\,\mathrm{kpc}$ fits cutouts are made for each channel, and the three channels are combined to make rgb thumbnails with \textsc{stiff}\footnote{https://www.astromatic.net/software/stiff} \citep{bertin2012}.
\par
To visually identify ram pressure (RP) candidates we follow the criteria given in \citet{roberts2020} for galaxies in the Coma Cluster.  These criteria were designed to identify galaxies which may be experiencing ram pressure stripping, however more generally they will flag strongly disturbed galaxies with asymmetric star formation.  In \citet{roberts2020} we show that, on average, galaxies matching these criteria are consistent with ram pressure stripping.  That said, for some galaxies, other cluster processes such as tidal effects or impulsive interactions (i.e.\ galaxy harassment) may certainly play a role.  We note that 13/17 of the Coma RP candidates identified by \citet{roberts2020} with stellar masses $>\!10^{9.5}\,\mathrm{M_\odot}$ (9/11 with stellar mass $>\!10^{10}\,\mathrm{M_\odot}$) have been subsequently observed to have clear ram pressure tails visible in the radio continuum with LOFAR at 144 MHz \citep{roberts2021_LOFARclust}. The LOFAR Two-metre Sky Survey suffers from substantial incompleteness for star-forming galaxies with $M_\mathrm{star} \lesssim 10^{9.5-10}\,\mathrm{M_\odot}$, therefore sensitivity limitations likely also play a role for those RP candidates from \citet{roberts2020} without detected radio continuum tails. The high recovery rate of radio continuum tails for Coma Cluster RP candidates reinforces that these visual criteria are relatively effective at identifying ram pressure stripping.  RP candidates are identified according to one (or more) of the following visual criteria:
\begin{enumerate}
    \itemsep0.5em
    \item \textit{The presence of asymmetric tails.} Observed on one side of the galaxy either in $u$-band (blue emission) or in dust (dark red extinction).

    \item \textit{Asymmetric star formation.} $u$-band (blue) emission that is knotty and clearly asymmetric about the galaxy centre.

    \item \textit{The presence of a bow shock morphology.}  $u$-band (blue emission) or dust (dark red extinction) concentrated along one edge of the galaxy, potentially tracing enhanced star formation along the leading edge at the galaxy-ICM interface.
\end{enumerate}
\noindent
 \citet{roberts2020} show that these visual criteria identify galaxies which are clearly disturbed according to quantitative morphology measures such as Asymmetry and $M_{20}$.  Here, Asymmetry is a measure of the rotational asymmetry and corresponds to the fractional galaxy flux that is contained in asymmetric components \citep[e.g.][]{abraham1996,conselice2003}, and $\mathrm{M_{20}}$ is a statistic derived from the second order moment brightest 20 per cent of a galaxy's flux that is particularly sensitive to bright features that are offset from galaxy centres \citep[e.g.][]{lotz2004}.
\par
We note that the tails referenced above should not be thought of as large scale extensions of material trailing the galaxy (for instance $\mathrm{H\alpha}$ or \textsc{Hi} ram pressure tails; e.g. \citealt{chung2009,poggianti2017,boselli2018}), but instead we use this `tail' notation to refer to filamentary or spiral arm like extensions of the galaxy that are asymmetric about the bright central component of the galaxy (i.e. these features are more extended towards one side of the galaxy than the other).  This is in contrast to the `asymmetric star formation' condition that is tracing asymmetric u-band emission (often knotty and compact) that could be embedded in an otherwise symmetric galaxy disk.  Shocks from ram pressure can induce enhanced gas densities in galaxies, in particular in a bow-like morphology along the leading edge of the galaxy \citep[e.g.][]{rasmussen2006,ramos-martinez2018,troncoso-iribarren2020}.  These enhanced densities in turn can lead to strong star formation along the leading edge \citep[e.g.][]{gavazzi2001,tomicic2018,boselli2021}.   It is the result of this ICM-ISM interaction that we are looking to identify when we consider `bow shock morphologies' in the u-band or dust.
\par
Visual inspections were completed by a single classifier (IDR).  We note that the exact criteria from \citet{roberts2020} were used, and in \citet{roberts2020} there was good agreement shown between multiple classifiers (e.g. 83 per cent of RP candidates were identified by at least 4/5 classifiers). Mergers are also flagged in this visual classification process, and galaxies identified as mergers are removed from the sample.  As in \citet{roberts2020}, mergers are flagged on the basis of two clearly interacting galaxies, or the presence of multiple, bright galaxy nuclei for the case of more evolved mergers.  This does not amount to an exhaustive merger identification, but it serves the purpose of ensuring that obvious mergers are not contaminating the RP candidate sample. Less than 2 per cent of classified galaxies were identified as mergers in this way.  Galaxies are classified in a randomized order and the coordinates of the galaxy and the ID of the host cluster are not known in the classification process.  48 RP candidates are identified by this process, which span the entire stellar mass range of the sample from $\sim\!10^{9}\,\mathrm{M_\odot}$ to $\sim\!10^{11}\,\mathrm{M_\odot}$.  29/48 RP candidates are hosted by group mass halos ($<\!10^{14}\,\mathrm{M_\odot}$) and 19/48 RP candidates are hosted by cluster mass halos ($\ge\!10^{14}\,\mathrm{M_\odot}$).  In Fig.~\ref{fig:RPexamples} an example of the $u$-band, $r$-band, and rgb imaging from CFIS is shown.  For comparison, we also show the $u$, $r$, and rgb SDSS images (with the same filters as the CFIS images).  The differences in resolution and depth is clear for both the $u$ and the $r$ band, highlighting the superior image quality of CFIS.  In Fig.~\ref{fig:RPexamples}, faint blue tails are barely visible in the SDSS image but clear in CFIS.  Furthermore, $\sim$arcsecond scale sources are resolved by the CFIS imaging, many of which are coincident with these tails, that are not seen at all in the SDSS imaging. In Appendix~\ref{sec:appendix_img} we show $u$, $r$, and rgb CFIS images for all of the RP candidates (Figs \ref{fig:RPimages1}-\ref{fig:RPimages3}) and we also include a table with the coordinates of all of the RP candidates as well as some brief notes on the morphological features used to classify each galaxy.
\par
As a test of our classification procedure, when visually classifying group galaxies we also randomly inject 500 galaxies from the field sample.  In Fig.~\ref{fig:RPexamples} we also show an example of the UNIONS and SDSS imaging for a randomly selected galaxy from the field sample.  These field galaxies are randomly injected and when classifying a galaxy the classifier does not know whether the galaxy is from the cluster or the field sample.  This is an important test of the process, as we are trying to flag galaxies according to cluster-specific processes (i.e.\ ram pressure stripping), field galaxies should not pass the classification criteria.  After all galaxies have been classified (group galaxies + 500 random field galaxies), we go back and check the false-positive rate of our classifications -- i.e. what fraction of galaxies from the field sample were classified as RP candidates.  We find that this false positive rate is 1.2 per cent, with 6/500 field galaxies passing the visual criteria for RP candidates.  All of the field galaxies flagged as RP candidates were selected on the basis of asymmetric star formation within the galaxy, no galaxies in the field sample were flagged as showing asymmetric tails of bow shock morphological features.  We note that this false positive rate is clearly below the fraction of RP candidates that we find among group galaxies (see Fig.~\ref{fig:RP_Mstar}), therefore our method appears reliable at morphologically identifying galaxies according to cluster-specific processes.
\par
As is shown in Fig.~\ref{fig:IQ_depth}, CFIS provides excellent image quality and relatively uniform depth across the diverse set of groups and clusters in our sample.  The methods used in this work are only possible thanks to this uniform imaging.  If there were large variations in image quality or depth across the groups and clusters in the sample, this could bias the selection RP candidates toward the systems with the highest quality imaging.  We note that we do not find any variations in the fraction of RP candidates identified as a function of image quality or depth, over the narrow range covered by the CFIS imaging.  Finally, we reiterate the gains provided by CFIS over other wide-area imaging surveys such as the SDSS.  Without the depth and resolution provided by CFIS it becomes very difficult to identify these morphological features with confidence (Fig.~\ref{fig:RPexamples}).  This is particularly true for faint, low-mass galaxies which are most strongly influenced by environment \citep[e.g.][]{haines2006,bamford2009,peng2010}.

\subsubsection*{A Note on Nomenclature:}

Throughout this paper we make comparisons between three main galaxy samples, and will use the following nomenclature to differentiate them.  ``Field galaxies'' refers to star-forming galaxies from the field sample described in Section~\ref{sec:field_sample}.  ``Group galaxies'' will refers to star-forming galaxies which are group/cluster members (as described in Section~\ref{sec:sdss_match}), but are not classified as RP candidates.  We refer to these galaxies as group galaxies as the majority of the systems in this sample have $M_h < 10^{14}\,\mathrm{M_\odot}$, that said the group galaxy sample does include some galaxies in halos with $M_h \ge 10^{14}\,\mathrm{M_\odot}$ which are typically considered clusters. ``RP candidates'' will refer to star-forming galaxies which were identified as ram pressure candidates according to the criteria described above.  Galaxies undergoing mergers have been visually identified and removed from all samples.  All galaxies considered in this paper are star-forming according to the criteria, $\mathrm{sSFR} > 10^{-11}\,\mathrm{yr^{-1}}$.
\par
Additionally, this work considers a sample of SDSS groups and clusters spanning two orders of magnitude in halo mass.  Throughout the paper we will generally refer to groups as systems with $M_h < 10^{14}\,\mathrm{M_\odot}$ and clusters as systems with $M_h \ge 10^{14}\,\mathrm{M_\odot}$.

\section{Results} \label{sec:results}

\subsection{Ram-pressure Candidate Demographics} \label{sec:RP_demographics}

\subsubsection{Stellar Mass} \label{sec:RP_Mstar}

\begin{figure}
    \centering
    \includegraphics[width=\columnwidth]{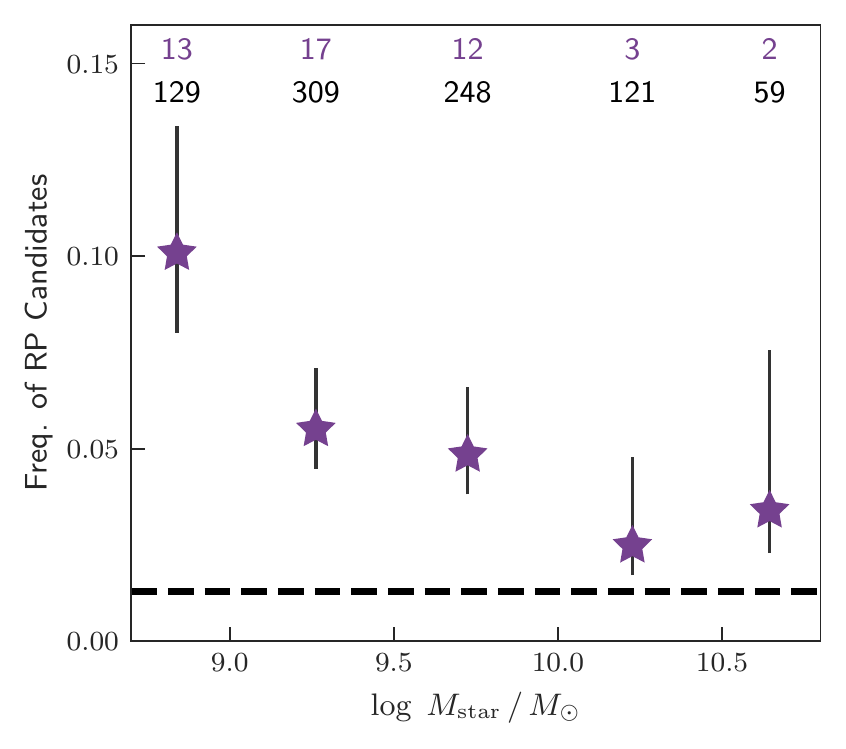}
    \caption{RP candidate fraction (see equation~\ref{eq:fRP}) as function of stellar mass.  Numbers along the top of the panel correspond to the number of RP candidates in each stellar mass bin (purple) and the total number of galaxies in each stellar mass bin (black).  The dashed black line shows the `false-positive rate' from our classifications, measured by randomly injecting field galaxies into the group galaxy classifications (see Section~\ref{sec:RPcandidates}).  Error bars correspond to 68 per cent binomial confidence intervals \citep{cameron2011}.}
    \label{fig:RP_Mstar}
\end{figure}

We first consider the frequency of RP candidates as a function of galaxy stellar mass.  We measure the frequency of RP candidates, which we define relative to all star-forming galaxies as
\begin{equation} \label{eq:fRP}
    f_\mathrm{RP} = \frac{N_\mathrm{RP}}{N_\mathrm{SF} + N_\mathrm{RP}}
\end{equation}
\noindent
where $N_\mathrm{RP}$ is the number of RP candidate galaxies, and $N_\mathrm{SF}$ is the number of normal star-forming galaxies (non-stripping).  $f_\mathrm{RP}$ gives the fraction of all star-forming galaxies in groups and clusters which have been identified as RP candidates.
\par
In Fig.~\ref{fig:RP_Mstar} we plot $f_\mathrm{RP}$ versus stellar mass for the sample of RP candidates, showing that $f_\mathrm{RP}$ increases from intermediate stellar masses to low stellar masses.  This is consistent with previous studies that argue low-mass galaxies are preferentially influenced by ram pressure stripping \citep[e.g.][]{fillingham2015,yun2019,roberts2019}, likely due to their shallow potential wells.  We note that the stellar mass distributions for galaxies in low-mass groups ($M_h < 10^{13.5}\,\mathrm{M_\odot}$), high-mass groups ($10^{13.5} \le M_h < 10^{14}\,\mathrm{M_\odot}$), and clusters ($M_h \ge 10^{14}\,\mathrm{M_\odot}$) are virtually identical.  Therefore the trend with stellar mass in Fig.~\ref{fig:RP_Mstar} is not due to different host halo masses for low- versus high-mass galaxies.

\subsubsection{Group/Cluster Mass} \label{sec:RP_Mh}

\begin{figure}
    \centering
    \includegraphics[width=\columnwidth]{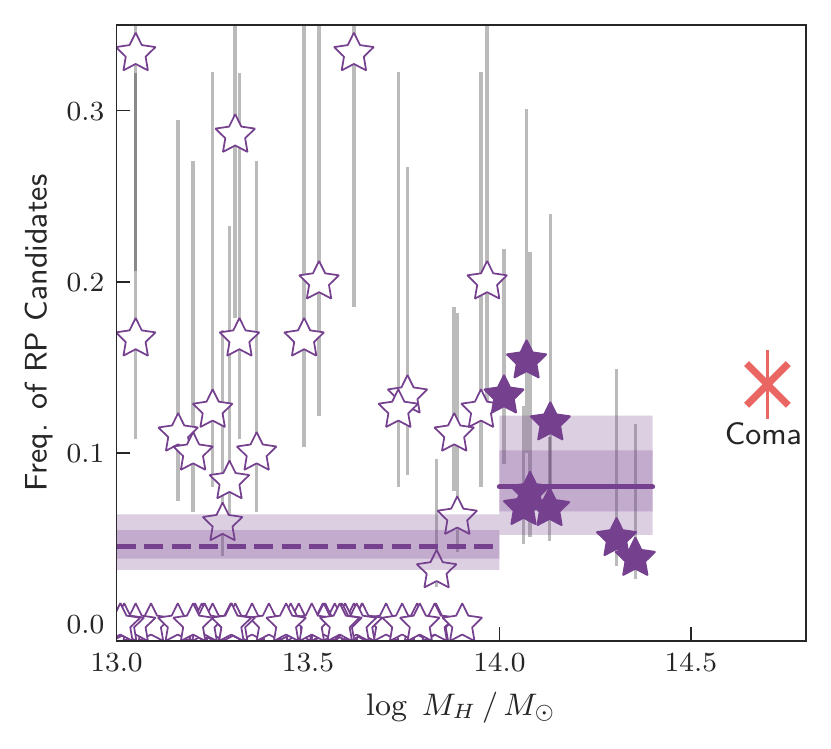}
    \caption{RP candidate fraction as a function of group/cluster halo mass.  Open stars correspond to individual galaxy groups ($M_H<10^{14}\,\mathrm{M_\odot}$) and filled stars correspond to individual galaxy clusters ($M_H<10^{14}\,\mathrm{M_\odot}$).  The horizontal dashed line shows the total fraction for all galaxy groups and the solid line shows the total fraction for all clusters.  Error bars are 68 per cent binomial confidence intervals and shaded regions correspond to 68 and 95 per cent confidence regions for the binned fractions \citep{cameron2011}.  We also show the ram pressure candidate frequency for galaxies in the Coma Cluster from \citet{roberts2020}.}
    \label{fig:RP_Mh}
\end{figure}

In Fig.~\ref{fig:RP_Mh} we plot the frequency of RP candidates as a function of group/cluster mass, ranging from low-mass groups ($\sim\!10^{13}\,\mathrm{M_\odot}$) to high-mass clusters ($\sim\!10^{14.5}\,\mathrm{M_\odot}$).  The data points mark $f_\mathrm{RP}$ for individual groups and clusters, and solid bands show the combined $f_\mathrm{RP}$ separately for groups ($M_h < 10^{14}\,\mathrm{M_\odot}$) and clusters ($M_h \ge 10^{14}\,\mathrm{M_\odot}$).  Open markers and the dashed line correspond to galaxy groups and filled markers and the solid line correspond to  galaxy clusters.  For reference, we also plot the frequency of RP candidates in the Coma Cluster presented in \citet{roberts2020}.
\par
On a system-by-system basis there is significant scatter in Fig.~\ref{fig:RP_Mh}, with the bulk of the scatter occurring at the low halo mass end.  The reason for the large scatter at low group masses is simply low-number statistics, as most groups in this sample host $<10$ star-forming members. The observed frequency of RP candidates in individual groups/clusters ranges from 0 to $\sim\!30$ per cent.  When considering the combined value of $f_\mathrm{RP}$ for groups and clusters separately, we find a small enhancement in the frequency of RP candidates for clusters compared to groups.  $f_\mathrm{RP}$ for clusters is larger than $f_\mathrm{RP}$ for groups by roughly a factor of two, and this difference is moderately significant at the $\sim\!2\sigma$ level.  Notably, all of systems with zero RP candidates fall in the group regime with $M_h < 10^{14}\,\mathrm{M_\odot}$, whereas all systems with $M_h \ge 10^{14}\,\mathrm{M_\odot}$ have detected RP candidates.  This may be a physical effect, in that the less dense ICM and smaller velocity dispersions in the group regime are less conducive to ram pressure stripping.  The high number of groups with no RP candidates could also be a result of the smaller galaxy memberships in lower mass halos.  Even if the intrinsic frequency of RP candidates does not depend on halo mass, there would naturally be more low-mass groups with zero RP candidates simply because they host fewer galaxies.

\begin{figure}
    \centering
    \includegraphics[width=\columnwidth]{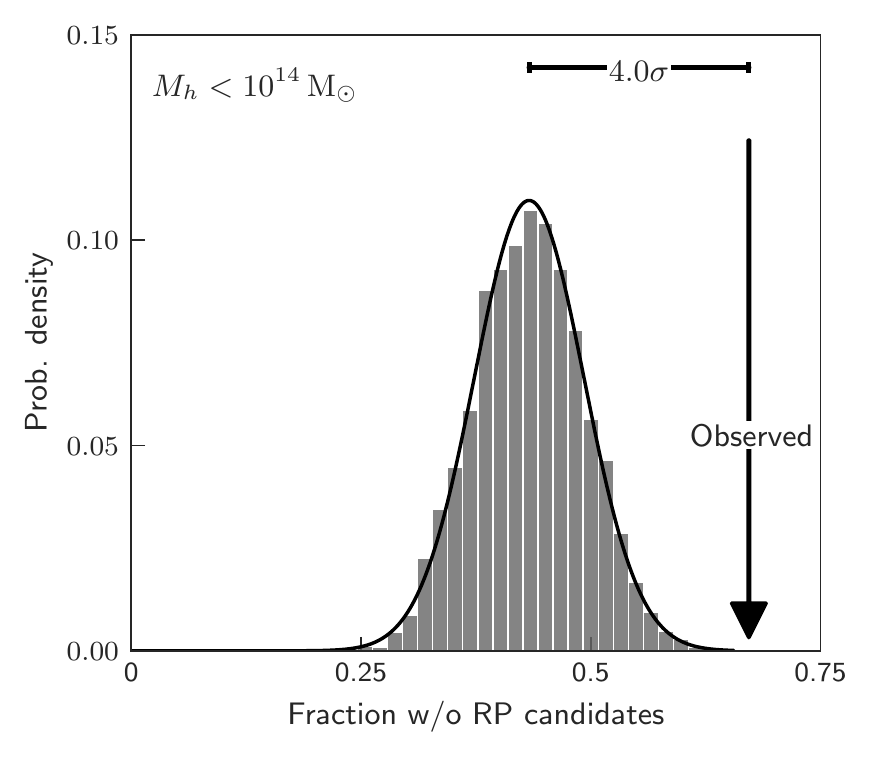}
    \caption{PDF corresponding to the fraction of groups ($M_h < 10^{14}\,\mathrm{M_\odot}$) expected to host zero ram pressure candidates, assuming the same RP frequency as clusters (8 per cent).  The observed fraction of groups with zero RP candidates (68 per cent) is also shown.}
    \label{fig:RPrand}
\end{figure}

Of the 61 groups in our sample with $M_h < 10^{14}\,\mathrm{M_\odot}$, we find that 41 (67 per cent) have zero star-forming galaxies classified as RP candidates.  We can now gauge whether this high fraction of systems with no RP candidates can be explained solely by the smaller galaxy memberships (as described above).  In other words, if we assume that galaxy groups ($M_h < 10^{14}\,\mathrm{M_\odot}$) and galaxy clusters ($M_h \ge 10^{14}\,\mathrm{M_\odot}$) have the same intrinsic frequency of RP candidates, and given the fact that they have fewer member galaxies, what fraction of groups would we expect to observe with zero RP candidates due to random chance alone?  To accomplish this we take the following procedure:
\begin{enumerate}
    \itemsep0.5em
    \item We take the observed frequency of RP candidates for clusters ($M_h \ge 10^{14}\,\mathrm{M_\odot}$) in our sample, which is 8 per cent.

    \item We take the observed number of star-forming galaxies in each group ($M_h < 10^{14}\,\mathrm{M_\odot}$), which ranges from 4 to 42 galaxies for our sample.

    \item We randomly draw a binary array ($0 \rightarrow$ normal star-forming galaxy, $1 \rightarrow$ RP candidate) for each group with length equal to the number of star-forming galaxies in that system.  The probability of drawing a 0 or 1 is weighted by 0.92 and 0.08 respectively, to reflect the frequency of RP candidates in clusters.

    \item For each group we count the number of non-zero elements in the random array, corresponding to the number of RP candidates expected in that system assuming the same frequency as clusters (8 per cent).  We then determine the fraction of groups expected to have zero RP candidates according to this random draw.

    \item We repeat this process for a total of 5000 iterations to generate a probability distribution function (PDF) for the fraction of groups which are expected to be observed with zero RP candidates.
\end{enumerate}
\noindent
The resulting PDF is shown in Fig.~\ref{fig:RPrand} along with the best-fit normal distribution, as well as the observed fraction of groups with zero RP candidates, of 68 per cent.  The observed fraction is clearly larger than expected given the assumption that the frequency of RP candidates does not depend on halo mass.  The significance of this discrepancy is $4\sigma$.  This implies that the large fraction of groups with zero RP candidates cannot be explained by galaxy membership alone, and instead the intrinsic frequency of RP candidates must be lower in groups than clusters.  In order to bring the observed fraction in Fig.~\ref{fig:RPrand} within the $1\sigma$ scatter of the PDF, one must assume that the frequency of RP candidates is smaller by a factor of two in groups compared to clusters.

\subsection{Projected Phase Space and $u$-band Asymmetry Orientations} \label{sec:PS}

\begin{figure}
    \centering
    \includegraphics[width=\columnwidth]{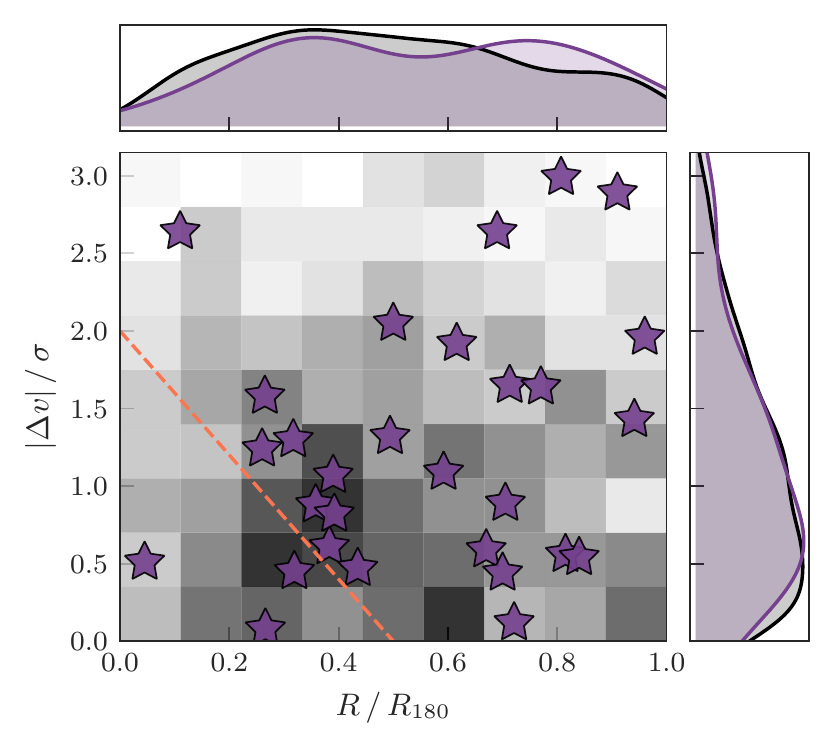}
    \caption{Projected phase space diagram.  The phase space distribution for group galaxies is shown with 2D histogram and ram pressure candidates are marked by purple stars.  The dashed line shows the virialized region (dominated by `ancient infallers') of phase space as given in \citet{rhee2017}.  We also show the marginal kernel density distributions for projected radius and velocity offset above the corresponding axes.}
    \label{fig:phase_space}
\end{figure}

In this section we explore the distribution of ram pressure candidates in projected phase space (velocity offset vs. projected radius) as well as estimate the direction of ram pressure based off of the orientation of the observed asymmetries in the $u$-band.

\subsubsection{Phase Space}

Previous works have shown that galaxies undergoing ram pressure stripping in clusters are systematically found at large velocity offsets in phase space \citep[e.g.][]{yoon2017,jaffe2018,roberts2020,roberts2021_LOFARclust}.  This corresponds to regions of phase space where ram pressure should be strong (since $P_\mathrm{ram} \sim \rho_\mathrm{ICM} v^2$), and also suggests that galaxies in clusters are primarily being stripped on their first infall \citep{jaffe2018}.  The majority ($\gtrsim\!60\%$) of RP candidates in this work are hosted by galaxy groups with $M_h < 10^{14}\,\mathrm{M_\odot}$, and the rest are primarily hosted by low-mass clusters ($M_h \sim 10^{14}\,\mathrm{M_\odot}$).  Therefore we can test whether the phase space trends reported for ram pressure stripping galaxies in clusters also hold in lower mass halos.  Unfortunately, we do not have sufficient statistics to do a group/cluster split from our sample alone.
\par
In Fig.~\ref{fig:phase_space} we show the distribution of group galaxies (greyscale 2D histogram) and RP candidates (purple stars) in projected phase space.  We also show the marginal distributions for normalized radius and velocity offsets above the corresponding axes.  Contrary to previous results for rich galaxy clusters, we find no difference between the phase space distribution for RP candidates and normal star-forming galaxies from this work.  There is no evidence in Fig.~\ref{fig:phase_space} that RP candidates are shifted to large velocity offsets or small group/cluster-centric radii.  This can be seen from the similarity of the marginal distributions for RP candidates (purple) and field galaxies (black) in Fig.~\ref{fig:phase_space}.  It is confirmed quantitatively with the two-sample Anderson-Darling test \citep{scholz1987}.  This could be a product of the weaker ram pressure, on average, in the lower mass halos that our sample probes, meaning that galaxies in groups may not show signs of stripping shortly after infall as is the case in clusters.  If galaxies in groups have ram pressure stripping features that persist beyond first infall, then that could explain the RP candidate distribution in Fig.~\ref{fig:phase_space} which appears well mixed and indistinguishable from the group galaxies.  This picture is supported by the work of \citet{oman2021} who model the timescales of gas stripping as a function of halo mass.  \citet{oman2021} find that galaxy groups ($M \sim 10^{13.5}\,\mathrm{M_\odot}$) strip their galaxies on longer timescales than clusters ($M \sim 10^{14.5}\,\mathrm{M_\odot}$), with a typical difference of $\sim\!3\,\mathrm{Gyr}$.  The distribution of RP candidates in Fig.~\ref{fig:phase_space} does not show strong evidence for galaxies (primarily in groups) being stripped primarily on first infall, as has been suggested for galaxies in clusters \citep[e.g.][]{yoon2017,jaffe2018,roberts2020}.

\subsubsection{$u$-band Asymmetry Orientations} \label{sec:orient}

\begin{figure}
    \centering
    \includegraphics[width=\columnwidth]{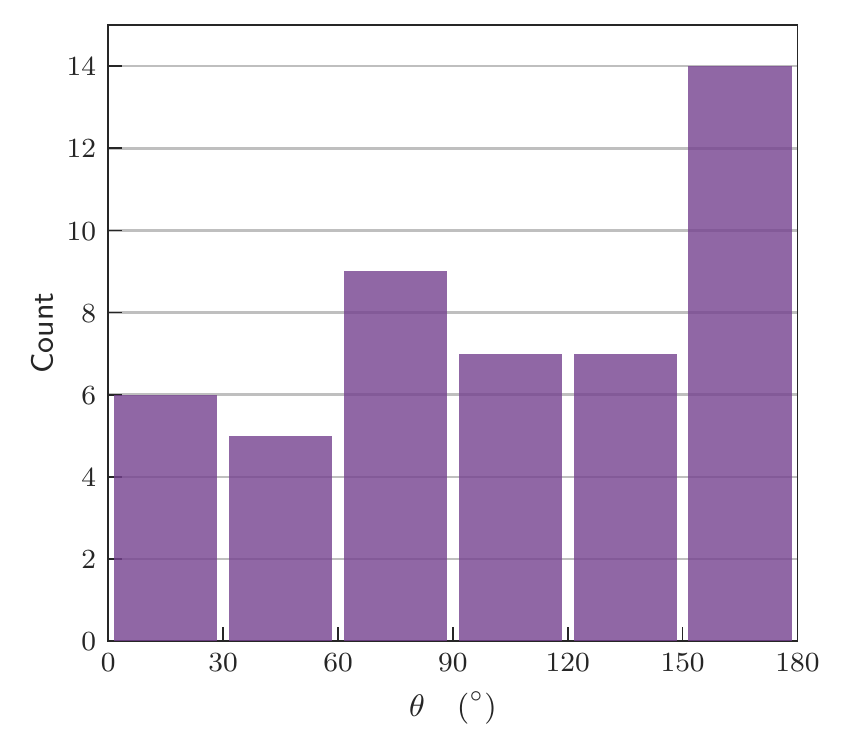}
    \caption{Orientation of $u$-band asymmetries for identified RP candidates.  An orientation of $0^\circ$ corresponds to an asymmetry directed toward the centre of the host group/cluster and an orientation of $180^\circ$ corresponds to an asymmetry oriented directly away from the centre of the host group/cluster.}
    \label{fig:orient}
\end{figure}

We now attempt to constrain the direction of ram pressure based on the observed asymmetries in RP candidates.  The distribution of these orientations contains information on orbital and infall history for galaxies undergoing ram pressure stripping; namely, galaxies stripped on first infall should have asymmetries oriented away from the group/cluster centre and galaxies being stripped after orbiting past pericentre should have asymmetries oriented toward the group/cluster centre.  While useful to gather information on galaxy orbits, this technique is also complicated by observational uncertainties.  Projection effects are a significant concern as observed orientations that are projected on the sky will not always be representative of the true 3D directions.  For this work specifically, there is added uncertainty given that we are observing ram pressure features in the rest-frame optical and therefore primarily tracing stellar distributions.  Ram pressure features are best traced by lower density components of the ISM (e.g.\ \textsc{Hi}, $\mathrm{H\alpha}$, radio continuum, etc.) that are more easily stripped by ram pressure.  Perturbations to the optical morphologies are more subtle, and therefore more ambiguous in terms of connecting to ram pressure.  That said, for most jellyfish galaxies in the Coma Cluster, observed radio continuum tails are roughly consistent with stripping directions estimated from optical imaging (see the appendices of \citealt{roberts2020} and \citealt{roberts2021_LOFARclust}).
\par
To estimate tail directions we use visual inspection of the colour images to assign the direction of the observed asymmetry for each galaxy an angle between $0^\circ$ and $360^\circ$, where $0^\circ = \mathrm{west}$ and $90^\circ = \mathrm{north}$. This is the same methodology used by \citet{roberts2020} for optically-selected RP candidates in Coma and by \citet{roberts2021_LOFARclust,roberts2021_LOFARgrp} for jellyfish galaxies in groups and clusters with radio continuum tails.  These directions are meant to be representative of the direction of ram pressure stripping; namely, if prominent, extended tails were present these are the directions along which we would expect to see them.  All of the estimated asymmetry directions are indicated on the panel images in Appendix~\ref{sec:appendix_img}.  From these asymmetry directions we then compute the angle with respect to the centre of the corresponding host group or cluster such that an orientation of $0^\circ$ corresponds to an asymmetry pointed directly at the group/cluster centre and an orientation of $180^\circ$ corresponds to an asymmetry pointed directly away from the group/cluster centre.  In Fig.~\ref{fig:orient} we show the distribution of asymmetry orientations for the RP candidates in this work.  Fig.~\ref{fig:orient} shows a relatively uniform distribution of angles, with the exception for the $[150^\circ,180^\circ]$ bin that shows evidence for an excess of RP candidates with observed asymmetries that are oriented away from the centre of the host group/cluster.  Given the uncertainties with regards to estimating orientations from rest-frame optical imaging (see previous paragraph), projection effects, and the relatively small number of RP candidates in our sample, it is difficult to make strong conclusions from Fig.~\ref{fig:orient}.  That said, the broad distribution of angles in Fig.~\ref{fig:orient} is consistent with a fairly mixed set of orbits amongst the RP candidates in this work, similar to the phase space results above. This is in contrast to asymmetry directions from optical imaging for RP candidates in the Coma Cluster which are almost exclusively above $120^\circ$ \citep{roberts2020}.  In Fig.~\ref{fig:orient} there is a small excess of orientations above $150^\circ$ which may indicate a slight overabundance of RP candidates on their first infall toward the group/cluster centre.

\subsection{Star Formation Properties} \label{sec:star_formation}

\begin{figure}
    \centering
    \includegraphics[width=\columnwidth]{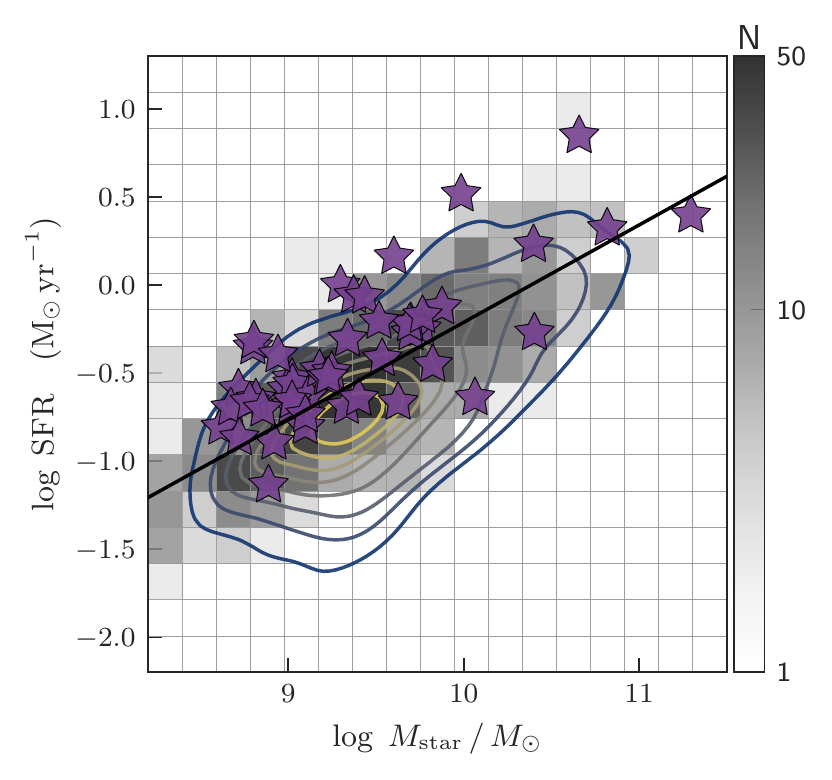}
    \caption{Star formation rate versus stellar mass for field galaxies (greyscale), group galaxies (contours), and ram pressure candidates (purple stars).  The solid line shows the best-fit star-forming main sequence relation for isolated field galaxies.}
    \label{fig:sfms}
\end{figure}

\begin{figure*}
    \centering
    \includegraphics[width=\textwidth]{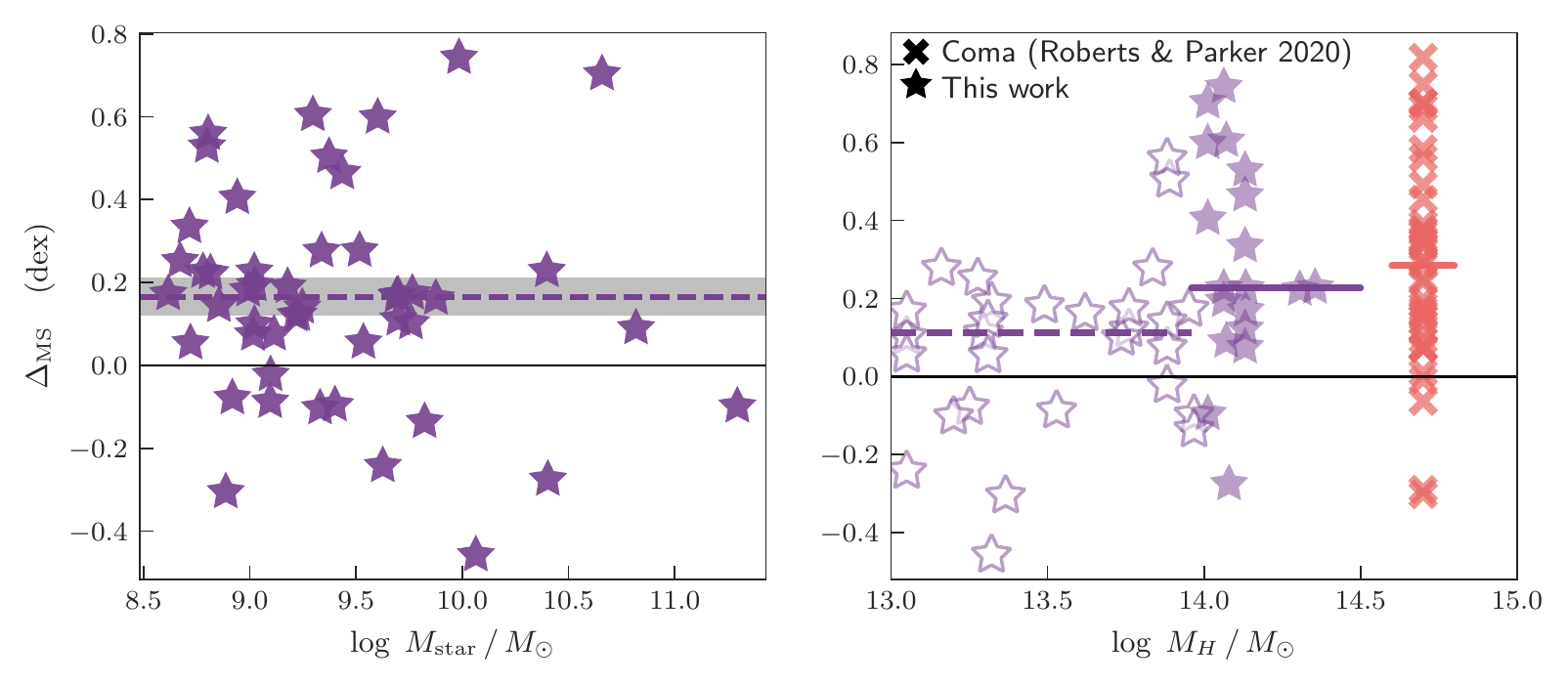}
    \caption{\textit{Left:}  Offset from the field star-forming main sequence as a function of stellar mass for ram pressure candidates.  The dashed line shows the median offset from the star-forming main sequence and the shaded region shows the standard error on the median. \textit{Right:}  Offset from the star-forming main sequence as a function of halo mass.  Open stars correspond to ram pressure candidates in groups, filled stars correspond to ram pressure candidates in clusters, and crosses are Coma Cluster ram pressure candidates from \citet{roberts2020}.  Horizontal bars show median offsets for galaxies in groups (dashed purple), clusters (solid purple), and Coma (solid orange).}
    \label{fig:deltaMS}
\end{figure*}

To probe cluster specific effects on galaxy star formation, we compare group galaxies to the isolated field sample.  We use the star-forming galaxies in the field sample to fit a field star-forming main sequence (SFMS) and compare cluster SFRs relative to this standard.  We consider all galaxies in the field sample with $\mathrm{sSFR} > 10^{-11}\,\mathrm{yr^{-1}}$ and fit a powerlaw to the $\mathrm{SFR}-M_\mathrm{star}$ relation.  The fit is done with the \textsc{linmix}\footnote{https://linmix.readthedocs.io/en/latest/index.html} \citep{kelly2007} package and accounts for uncertainties on both $\mathrm{SFR}$ and $M_\mathrm{star}$ for each field galaxy.  The best-fit SFMS relation for the field sample is $\log\mathrm{SFR} = 0.55 \times \log M_\mathrm{star} - 5.7$.
\par
In Fig.~\ref{fig:sfms} we plot SFR versus stellar mass.  Star-forming field galaxies are shown by the greyscale 2D histogram, the distribution of star-forming group galaxies are shown by the contours, and RP candidates are shown by the purple stars.  The solid black line marks the best fit SFMS for the field sample.  It is clear from Fig.~\ref{fig:sfms} that RP candidates are systematically found above the field SFMS, quantitatively, 77 per cent of RP candidates lie above the field SFMS.  This enhanced star formation in RP candidates has been reported by \citet{roberts2020} for galaxies in the Coma Cluster, and here is confirmed for RP candidates across a wide range in cluster mass.
\par
For all RP candidates, we now compute the offset in SFR relative to the field SFMS, $\Delta_\mathrm{MS}$.  This allows us to quantify the enhancement of star formation for RP candidates as a function of stellar and halo mass.  In Fig.~\ref{fig:deltaMS} we plot the offset from the SFMS as a function of stellar mass (left) and halo mass (right).  Data points show the offsets for individual RP candidates.  An enhancement in star formation ($\Delta_\mathrm{MS}>0$) is present across the entire range in stellar mass, except, perhaps, the high-mass end where the number of RP candidates is low.  The typical value for $\Delta_\mathrm{MS}$ is  $\sim\! 0.2\,\mathrm{dex}$ and there is no clear trend between $\Delta_\mathrm{MS}$ and stellar mass.
\par
In Fig.~\ref{fig:deltaMS} (right) we show $\Delta_\mathrm{MS}$ as a function of halo mass, with RP candidates in groups marked by the open symbols and RP candidates in clusters marked by the filled symbols.  We also show $\Delta_\mathrm{MS}$ for Coma RP candidates from \citet{roberts2020} at the high halo mass end.  The median $\Delta_\mathrm{MS}$ for RP candidates in groups, clusters, and Coma are shown by the horizontal lines. A mild difference is seen in Fig.~\ref{fig:deltaMS} with a larger median $\Delta_\mathrm{MS}$ in more massive halos.  RP candidates hosted by clusters in this sample as well as the Coma Cluster \citep{roberts2020} have larger offsets from the SFMS than RP candidates hosted by groups.  Furthermore, RP candidates in clusters scatter to much larger $\Delta_\mathrm{MS}$ values than is seen for group RP candidates.  The RP candidates found far above the SFMS in Fig.~\ref{fig:sfms} are largely hosted by relatively massive clusters, which may be connected to stronger ram pressure in these systems.  We reiterate that all offsets from the SFMS are measured relative to the field SFMS.  When compared to normal star-forming galaxies in clusters, RP candidates show enhanced star formation at all stellar masses and within clusters of all halo masses.

\section{Discussion} \label{sec:discussion}

\citet{roberts2020} show that the visual criteria listed in Section~\ref{sec:RPcandidates} select a population of galaxies which are consistent with ram pressure stripping as the primary driver behind the observed disturbed morphologies.  Furthermore, many of the Coma RP candidates identified by \citet{roberts2020} have been subsequently observed to have clear ram pressure tails visible in the radio continuum with LOFAR at 144 MHz \citep{roberts2021_LOFARclust}, reinforcing that these visual criteria are effective at identifying ram pressure stripping. In this work, the stellar-mass dependence of RP candidates (i.e.\ preferentially low-mass galaxies) and the group/cluster mass dependence of RP candidates (i.e.\ preferentially found in higher mass clusters) are measured and are consistent with simple predictions from models of ram pressure.  That said, it is possible that some galaxies which we identify as RP candidates may owe their disturbed morphologies (at least in part) to other physical processes, for example tidal effects or galaxy harassment.  Moving forward, more multiwavelength data on these galaxies, such as $\mathrm{H\alpha}$ or \textsc{Hi}, would be an asset to more definitively constrain the origins their observed morphologies. The LOFAR Two-metre Sky Survey \citep[LoTSS;][]{shimwell2017,shimwell2019} will finish imaging the entire northern extragalactic sky at $144\,\mathrm{MHz}$ within the next few years, therefore we will be able to measure how many of these RP candidates show extended radio continuum tails. Though the LoTSS sensitivity may not be high enough to detect the lowest mass RP candidates at $144\,\mathrm{MHz}$ \citep{roberts2021_LOFARclust}.
\par
Given that the criteria in Section~\ref{sec:RPcandidates} were designed to identify ram pressure stripping galaxies, and that the results from this work and \citet{roberts2020} are consistent with expectations from ram pressure stripping, for the remained of the discussion we will assume that the morphologies of RP candidates in this work are primarily being driven by ram pressure.

\subsection{Stellar Mass Dependence and Star Formation}

In Fig.~\ref{fig:RP_Mstar} we find a clear mass dependence for the frequency of RP candidates.  The highest values for $f_\mathrm{RP}$ correspond to the lowest mass galaxies in the sample.  This increase in $f_\mathrm{RP}$ toward the lowest masses is likely to be physical in nature.  It has been previously shown that environmental quenching is most important for low-mass galaxies ($M_\mathrm{star} \lesssim \mathrm{few} \times 10^{10}\,\mathrm{M_\odot}$, \citealt{haines2006,bamford2009}), therefore there is some expectation that low-mass galaxies should be strongly affected by environmental processes.  Given the relatively shallow potential wells of low-mass galaxies, they may be more susceptible to environmental processes such as ram pressure stripping.  This is consistent with the observed trend in Fig.~\ref{fig:RP_Mstar}.  \citet{yun2019} visually identify jellyfish galaxies (according the gas column density maps) in the Illustris TNG simulation, finding that low-mass galaxies are most frequently jellyfish, and the fraction of jellyfish galaxies (relative to all gas-rich galaxies) decreases monotonically with increasing galaxy stellar mass.  Other observational studies have also argued that low-mass galaxies are preferentially affected by ram pressure stripping relative to higher-mass counterparts \citep{fillingham2015,wetzel2015,rodriguez-wimberly2019,roberts2019,durret2021}.
\par
Across the stellar mass range we see evidence for enhanced star formation (Figs \ref{fig:sfms} \& \ref{fig:deltaMS}) for RP candidates.  The star formation enhancement is clear through the SED SFRs \citep{salim2016,salim2018} used in the text.  We also confirm that SDSS fibre $\mathrm{H\alpha}$ fluxes are strongly enhanced in RP candidates, which trace only the central $3''$ of each galaxy.  In a ram pressure stripping context, the enhanced SFRs may be a result of compression and increased gas densities in the disc due to the galaxy-ICM interaction.  When comparing RP candidates to the field SFMS, we find enhancements of roughly 0.2 dex, with no strong dependence on stellar mass.  The enhanced SFRs in ram pressure stripping galaxies are consistent with previous work from observations \citep{dressler1983,ebeling2014,poggianti2016,vulcani2018b,roberts2020,wang2020,durret2021} and simulations \citep{steinhauser2012,bekki2014,troncoso-iribarren2016}.  In Fig.~\ref{fig:deltaMS} (right) we see evidence for a weak dependence of SFMS offset on host halo mass.  The observed offsets are slightly larger for RP candidates hosted by clusters ($M_h \ge 10^{14}\,\mathrm{M_\odot}$) compared to groups ($M_h < 10^{14}\,\mathrm{M_\odot}$), and there is significant scatter toward large main sequence offsets for group galaxies.  If robust, this trend could be related to stronger ram pressure in clusters, leading to more extreme gas compression which catalyzes stronger bursts of star formation.  This can potentially be tested observationally with recent and future surveys of molecular gas in group galaxies (e.g. VERTICO\footnote{https://sites.google.com/view/verticosurvey/home}, the ALMA Fornax Cluster Survey, GASP; Brown et al. in prep, \citealt{zabel2019}, \citealt{moretti2020}). It is also important to note that we identify a relatively small number of RP candidates in the group regime, therefore it is difficult to make strong conclusions regarding offsets from the SFMS in groups without a larger sample.

\subsection{Low-Mass Groups to Massive Clusters}

Galaxies in our RP candidate sample span host halo masses ranging from $\sim\!10^{13}-10^{14.5}\,\mathrm{M_\odot}$, and even up to $\sim\!10^{15}\,\mathrm{M_\odot}$ when including Coma Cluster RP candidates from \citet{roberts2020}.  The fact that we observe RP candidates in the group regime is consistent with previous works that have shown evidence for ram pressure stripping in galaxy groups \citep{rasmussen2006,vulcani2018a}.  Ram pressure scales as $\rho_\mathrm{ICM}v^2$ \citep[e.g.][]{gunn1972}, and both the density of the ICM and the relative velocities of satellite galaxies increase with cluster mass, at least on average.  Therefore galaxies should experience stronger ram pressure in high-mass clusters than low-mass groups.  We do find evidence for more RP candidates in clusters versus groups by roughly a factor of two (see Figs~\ref{fig:RP_Mh} \& \ref{fig:RPrand}).  This is consistent with simulation results from \citet{yun2019} that show a factor of $\sim$2 increase in the jellyfish fraction from low-mass groups ($M_h \sim 10^{13}\,\mathrm{M_\odot}$) to galaxy clusters ($M_h > 10^{14}\,\mathrm{M_\odot}$).  It is difficult to directly compare the jellyfish fraction in \citet{yun2019} to the RP candidate frequencies in this work, due to the very different methodologies (i.e. jellyfish galaxies identified from simulated gas column densities vs. RP candidates from rest-frame optical imaging).  That said the general trends with stellar mass and halo mass from \citeauthor{yun2019} are consistent with the observed trends in this work.  Most of the groups in our sample have a relatively small number of spectroscopic members, with a median number of $\sim$10 star-forming members.  Therefore on an individual group-by-group basis the frequency of RP candidates is highly susceptible to low-number statistics, as is apparent from the scatter in Fig~\ref{fig:RP_Mh}.
\par
Given that ram pressure is expected to be weaker in the group regime, galaxies in groups retain a larger fraction of their gas over the course of an orbit, whereas galaxies in clusters should be stripped more quickly.  Galaxies in groups may  stripped more slowly relative to galaxies in clusters.  This picture is also consistent with the quenching models in \citet{oman2021}.  If ram pressure stripping features survive for longer in galaxy groups this would increase the number of RP candidates identified in groups relative to clusters, contributing to the mild dependence of RP candidate frequency on halo mass seen in this work.  The phase space distribution in Fig.~\ref{fig:phase_space} suggests that RP candidates in groups persist beyond first infall, consistent with this picture.

\section{Conclusions} \label{sec:conclusion}

In this work we present a sample of ram pressure RP candidate galaxies, visually identified from the Canada-France Imaging Survey, which covers $>\!50$ SDSS groups and clusters ranging in mass from $\sim\!10^{13}-10^{14.5}\,\mathrm{M_\odot}$.  This large sample allows us to study the properties and demographics of galaxies likely being influenced by ram pressure in nearby groups and clusters.  The primary findings of this work are as follows:

\begin{enumerate}
    \itemsep0.5em

    \item The frequency of ram pressure candidates is highest for low-mass galaxies.

    \item The average frequency of ram pressure candidates is larger in clusters ($M_h \ge 10^{14}\,\mathrm{M_\odot}$) than groups ($M_h < 10^{14}\,\mathrm{M_\odot}$) by roughly a factor of two.  On an individual group/cluster basis, the frequency of ram pressure candidates shows significant scatter with halo mass.

    \item Ram pressure candidates have projected phase space distributions which are consistent with normal star-forming group galaxies.  Since the majority of ram pressure candidates in this work are hosted by galaxy groups, this suggests that galaxies in groups may show signature of ram pressure stripping well after first infall.

    \item Ram pressure candidates show a broad distribution of observed $u$-band asymmetry orientations, however there is a small excess of ram pressure candidates with asymmetric features oriented directly away from the group/cluster centre.

    \item Ram pressure candidates have enhanced star formation rates, relative to both isolated field galaxies and normal star-forming group galaxies.  Median offsets from the star-forming main sequence for RP candidates do not depend strongly on stellar mass, but are slightly larger for ram pressure candidates in clusters compared to groups.
\end{enumerate}

In this paper we have presented an introductory look at this sample of CFIS ram pressure candidates and the properties of these disturbed galaxies.  In the next paper in this series we will move beyond visual classifications and explore the quantitative morphological properties (e.g. Gini-$\mathrm{M_{20}}$, concentration-asymmetry) of the ram pressure candidates identified in this work.  Ultimately looking to determine whether an automated selection based on quantitative morphology measures can reproduce the sample of visually identified ram pressure candidates from this work.  This will be especially important in the context of next generation photometric galaxy surveys (e.g. EUCLID, the Vera C. Rubin Observatory, the Roman Space Telescope), where the number of galaxies with high-quality imaging will surpass what is practical for by-eye classifications.  The large number of ram pressure candidates, across a range in redshifts, that could be identified with these upcoming wide-field imaging projects will significantly advance our understanding of the prevalence and importance of ram pressure stripping in dense environments.

\section*{Acknowledgments}

IDR acknowledges support from the ERCStarting  Grant  Cluster  Web  804208.  LCP and MJH acknowledge funding from the Natural Sciences and Engineering Research Council of Canada.
\par
This work is based on data obtained as part of the Canada-France Imaging Survey, a CFHT large program of the National Research Council of Canada and the French Centre National de la Recherche Scientifique. Based on observations obtained with MegaPrime/MegaCam, a joint project of CFHT and CEA Saclay, at the Canada-France-Hawaii Telescope (CFHT) which is operated by the National Research Council (NRC) of Canada, the Institut National des Science de l'Univers (INSU) of the Centre National de la Recherche Scientifique (CNRS) of France, and the University of Hawaii.

\section*{Data Availability}

The SDSS catalogs used in this work are available at https://salims.pages.iu.edu/gswlc/ and https://gax.sjtu.edu.cn/data/Group.html.  A subset of the raw imaging data underlying this article are publicly available via the Canadian Astronomical Data Center at http://www.cadc-ccda.hia-iha.nrc-cnrc.gc.ca/en/megapipe/.  The remaining raw data and all processed data are available to members of the Canadian and French communities via reasonable requests to the principal investigators of the Canada-France Imaging Survey, Alan McConnachie and Jean-Charles Cuillandre. All data will be publicly available to the international community at the end of the proprietary period, scheduled for 2023.




\bibliographystyle{mnras}
\bibliography{mnras_template} 

\begin{thebibliography}{}
\makeatletter
\relax
\def\mn@urlcharsother{\let\do\@makeother \do\$\do\&\do\#\do\^\do\_\do\%\do\~}
\def\mn@doi{\begingroup\mn@urlcharsother \@ifnextchar [ {\mn@doi@}
  {\mn@doi@[]}}
\def\mn@doi@[#1]#2{\def\@tempa{#1}\ifx\@tempa\@empty \href
  {http://dx.doi.org/#2} {doi:#2}\else \href {http://dx.doi.org/#2} {#1}\fi
  \endgroup}
\def\mn@eprint#1#2{\mn@eprint@#1:#2::\@nil}
\def\mn@eprint@arXiv#1{\href {http://arxiv.org/abs/#1} {{\tt arXiv:#1}}}
\def\mn@eprint@dblp#1{\href {http://dblp.uni-trier.de/rec/bibtex/#1.xml}
  {dblp:#1}}
\def\mn@eprint@#1:#2:#3:#4\@nil{\def\@tempa {#1}\def\@tempb {#2}\def\@tempc
  {#3}\ifx \@tempc \@empty \let \@tempc \@tempb \let \@tempb \@tempa \fi \ifx
  \@tempb \@empty \def\@tempb {arXiv}\fi \@ifundefined
  {mn@eprint@\@tempb}{\@tempb:\@tempc}{\expandafter \expandafter \csname
  mn@eprint@\@tempb\endcsname \expandafter{\@tempc}}}

\bibitem[\protect\citeauthoryear{{Abraham}, {Tanvir}, {Santiago}, {Ellis},
  {Glazebrook}  \& {van den Bergh}}{{Abraham} et~al.}{1996}]{abraham1996}
{Abraham} R.~G.,  {Tanvir} N.~R.,  {Santiago} B.~X.,  {Ellis} R.~S.,
  {Glazebrook} K.,   {van den Bergh} S.,  1996, \mn@doi [MNRAS]
  {10.1093/mnras/279.3.L47}, \href
  {https://ui.adsabs.harvard.edu/abs/1996MNRAS.279L..47A} {279, L47}

\bibitem[\protect\citeauthoryear{{Bamford} et~al.,}{{Bamford}
  et~al.}{2009}]{bamford2009}
{Bamford} S.~P.,  et~al., 2009, \mn@doi [MNRAS]
  {10.1111/j.1365-2966.2008.14252.x}, \href
  {http://adsabs.harvard.edu/abs/2009MNRAS.393.1324B} {393, 1324}

\bibitem[\protect\citeauthoryear{{Bekki}}{{Bekki}}{2014}]{bekki2014}
{Bekki} K.,  2014, \mn@doi [MNRAS] {10.1093/mnras/stt2216}, \href
  {https://ui.adsabs.harvard.edu/abs/2014MNRAS.438..444B} {438, 444}

\bibitem[\protect\citeauthoryear{{Bertin}}{{Bertin}}{2012}]{bertin2012}
{Bertin} E.,  2012, {Displaying Digital Deep Sky Images}.
p.~263

\bibitem[\protect\citeauthoryear{{Boissier} et~al.,}{{Boissier}
  et~al.}{2012}]{boissier2012}
{Boissier} S.,  et~al., 2012, \mn@doi [A\&A] {10.1051/0004-6361/201219957},
  \href {https://ui.adsabs.harvard.edu/abs/2012A&A...545A.142B} {545, A142}

\bibitem[\protect\citeauthoryear{{Boselli} et~al.,}{{Boselli}
  et~al.}{2018}]{boselli2018}
{Boselli} A.,  et~al., 2018, \mn@doi [A\&A] {10.1051/0004-6361/201732407},
  \href {https://ui.adsabs.harvard.edu/abs/2018A&A...614A..56B} {614, A56}

\bibitem[\protect\citeauthoryear{{Boselli} et~al.,}{{Boselli}
  et~al.}{2021}]{boselli2021}
{Boselli} A.,  et~al., 2021, \mn@doi [A\&A] {10.1051/0004-6361/202039046},
  \href {https://ui.adsabs.harvard.edu/abs/2021A&A...646A.139B} {646, A139}

\bibitem[\protect\citeauthoryear{{Brown} et~al.,}{{Brown}
  et~al.}{2017}]{brown2017}
{Brown} T.,  et~al., 2017, \mn@doi [MNRAS] {10.1093/mnras/stw2991}, \href
  {https://ui.adsabs.harvard.edu/abs/2017MNRAS.466.1275B} {466, 1275}

\bibitem[\protect\citeauthoryear{{Cameron}}{{Cameron}}{2011}]{cameron2011}
{Cameron} E.,  2011, \mn@doi [PASA] {10.1071/AS10046}, \href
  {http://adsabs.harvard.edu/abs/2011PASA...28..128C} {28, 128}

\bibitem[\protect\citeauthoryear{{Chambers} et~al.,}{{Chambers}
  et~al.}{2016}]{chambers2016}
{Chambers} K.~C.,  et~al., 2016, arXiv e-prints, \href
  {https://ui.adsabs.harvard.edu/abs/2016arXiv161205560C} {p. arXiv:1612.05560}

\bibitem[\protect\citeauthoryear{{Chen} et~al.,}{{Chen}
  et~al.}{2020}]{chen2020}
{Chen} H.,  et~al., 2020, \mn@doi [MNRAS] {10.1093/mnras/staa1868}, \href
  {https://ui.adsabs.harvard.edu/abs/2020MNRAS.496.4654C} {496, 4654}

\bibitem[\protect\citeauthoryear{{Chung}, {van Gorkom}, {Kenney}  \&
  {Vollmer}}{{Chung} et~al.}{2007}]{chung2007}
{Chung} A.,  {van Gorkom} J.~H.,  {Kenney} J.~D.~P.,   {Vollmer} B.,  2007,
  \mn@doi [ApJl] {10.1086/518034}, \href
  {http://adsabs.harvard.edu/abs/2007ApJ...659L.115C} {659, L115}

\bibitem[\protect\citeauthoryear{{Chung}, {van Gorkom}, {Kenney}, {Crowl}  \&
  {Vollmer}}{{Chung} et~al.}{2009}]{chung2009}
{Chung} A.,  {van Gorkom} J.~H.,  {Kenney} J.~D.~P.,  {Crowl} H.,   {Vollmer}
  B.,  2009, \mn@doi [AJ] {10.1088/0004-6256/138/6/1741}, \href
  {http://adsabs.harvard.edu/abs/2009AJ....138.1741C} {138, 1741}

\bibitem[\protect\citeauthoryear{{Clemens}, {Alexander}  \& {Green}}{{Clemens}
  et~al.}{2000}]{clemens2000}
{Clemens} M.~S.,  {Alexander} P.,   {Green} D.~A.,  2000, \mn@doi [MNRAS]
  {10.1046/j.1365-8711.2000.03069.x}, \href
  {https://ui.adsabs.harvard.edu/abs/2000MNRAS.312..236C} {312, 236}

\bibitem[\protect\citeauthoryear{{Conselice}}{{Conselice}}{2003}]{conselice2003}
{Conselice} C.~J.,  2003, \mn@doi [ApJS] {10.1086/375001}, \href
  {https://ui.adsabs.harvard.edu/abs/2003ApJS..147....1C} {147, 1}

\bibitem[\protect\citeauthoryear{{Crowl}, {Kenney}, {van Gorkom}  \&
  {Vollmer}}{{Crowl} et~al.}{2005}]{crowl2005}
{Crowl} H.~H.,  {Kenney} J. D.~P.,  {van Gorkom} J.~H.,   {Vollmer} B.,  2005,
  \mn@doi [AJ] {10.1086/430526}, \href
  {https://ui.adsabs.harvard.edu/abs/2005AJ....130...65C} {130, 65}

\bibitem[\protect\citeauthoryear{{Darg} et~al.,}{{Darg}
  et~al.}{2010}]{darg2010}
{Darg} D.~W.,  et~al., 2010, \mn@doi [MNRAS]
  {10.1111/j.1365-2966.2009.15786.x}, \href
  {https://ui.adsabs.harvard.edu/abs/2010MNRAS.401.1552D} {401, 1552}

\bibitem[\protect\citeauthoryear{{Dressler}}{{Dressler}}{1980}]{dressler1980}
{Dressler} A.,  1980, \mn@doi [ApJ] {10.1086/157753}, \href
  {http://adsabs.harvard.edu/abs/1980ApJ...236..351D} {236, 351}

\bibitem[\protect\citeauthoryear{{Dressler} \& {Gunn}}{{Dressler} \&
  {Gunn}}{1983}]{dressler1983}
{Dressler} A.,  {Gunn} J.~E.,  1983, \mn@doi [ApJ] {10.1086/161093}, \href
  {https://ui.adsabs.harvard.edu/abs/1983ApJ...270....7D} {270, 7}

\bibitem[\protect\citeauthoryear{{Durret}, {Chiche}, {Lobo}  \&
  {Jauzac}}{{Durret} et~al.}{2021}]{durret2021}
{Durret} F.,  {Chiche} S.,  {Lobo} C.,   {Jauzac} M.,  2021, arXiv e-prints,
  \href {https://ui.adsabs.harvard.edu/abs/2021arXiv210202595D} {p.
  arXiv:2102.02595}

\bibitem[\protect\citeauthoryear{{Ebeling}, {Stephenson}  \& {Edge}}{{Ebeling}
  et~al.}{2014}]{ebeling2014}
{Ebeling} H.,  {Stephenson} L.~N.,   {Edge} A.~C.,  2014, \mn@doi [ApJL]
  {10.1088/2041-8205/781/2/L40}, \href
  {https://ui.adsabs.harvard.edu/abs/2014ApJ...781L..40E} {781, L40}

\bibitem[\protect\citeauthoryear{{Fillingham}, {Cooper}, {Wheeler},
  {Garrison-Kimmel}, {Boylan-Kolchin}  \& {Bullock}}{{Fillingham}
  et~al.}{2015}]{fillingham2015}
{Fillingham} S.~P.,  {Cooper} M.~C.,  {Wheeler} C.,  {Garrison-Kimmel} S.,
  {Boylan-Kolchin} M.,   {Bullock} J.~S.,  2015, \mn@doi [MNRAS]
  {10.1093/mnras/stv2058}, \href
  {http://adsabs.harvard.edu/abs/2015MNRAS.454.2039F} {454, 2039}

\bibitem[\protect\citeauthoryear{{Gaia Collaboration} et~al.,}{{Gaia
  Collaboration} et~al.}{2016}]{gaia2016}
{Gaia Collaboration} et~al., 2016, \mn@doi [A\&A]
  {10.1051/0004-6361/201629272}, \href
  {https://ui.adsabs.harvard.edu/abs/2016A&A...595A...1G} {595, A1}

\bibitem[\protect\citeauthoryear{{Gaia Collaboration} et~al.,}{{Gaia
  Collaboration} et~al.}{2018}]{gaiaDR2}
{Gaia Collaboration} et~al., 2018, \mn@doi [A\&A]
  {10.1051/0004-6361/201833051}, \href
  {https://ui.adsabs.harvard.edu/abs/2018A&A...616A...1G} {616, A1}

\bibitem[\protect\citeauthoryear{{Gavazzi} \& {Jaffe}}{{Gavazzi} \&
  {Jaffe}}{1987}]{gavazzi1987}
{Gavazzi} G.,  {Jaffe} W.,  1987, A\&A, \href
  {https://ui.adsabs.harvard.edu/abs/1987A&A...186L...1G} {186, L1}

\bibitem[\protect\citeauthoryear{{Gavazzi}, {Boselli}, {Mayer},
  {Iglesias-Paramo}, {V{\'\i}lchez}  \& {Carrasco}}{{Gavazzi}
  et~al.}{2001}]{gavazzi2001}
{Gavazzi} G.,  {Boselli} A.,  {Mayer} L.,  {Iglesias-Paramo} J.,
  {V{\'\i}lchez} J.~M.,   {Carrasco} L.,  2001, \mn@doi [ApJL]
  {10.1086/338389}, \href
  {https://ui.adsabs.harvard.edu/abs/2001ApJ...563L..23G} {563, L23}

\bibitem[\protect\citeauthoryear{{George} et~al.,}{{George}
  et~al.}{2018}]{george2018}
{George} K.,  et~al., 2018, \mn@doi [MNRAS] {10.1093/mnras/sty1452}, \href
  {https://ui.adsabs.harvard.edu/abs/2018MNRAS.479.4126G} {479, 4126}

\bibitem[\protect\citeauthoryear{{Gunn} \& {Gott}}{{Gunn} \&
  {Gott}}{1972}]{gunn1972}
{Gunn} J.~E.,  {Gott} III J.~R.,  1972, \mn@doi [ApJ] {10.1086/151605}, \href
  {http://adsabs.harvard.edu/abs/1972ApJ...176....1G} {176, 1}

\bibitem[\protect\citeauthoryear{{Haines}, {La Barbera}, {Mercurio}, {Merluzzi}
   \& {Busarello}}{{Haines} et~al.}{2006}]{haines2006}
{Haines} C.~P.,  {La Barbera} F.,  {Mercurio} A.,  {Merluzzi} P.,   {Busarello}
  G.,  2006, \mn@doi [ApJL] {10.1086/507297}, \href
  {http://adsabs.harvard.edu/abs/2006ApJ...647L..21H} {647, L21}

\bibitem[\protect\citeauthoryear{{Haines} et~al.,}{{Haines}
  et~al.}{2015}]{haines2015}
{Haines} C.~P.,  et~al., 2015, \mn@doi [ApJ] {10.1088/0004-637X/806/1/101},
  \href {http://adsabs.harvard.edu/abs/2015ApJ...806..101H} {806, 101}

\bibitem[\protect\citeauthoryear{{Hester}}{{Hester}}{2006}]{hester2006}
{Hester} J.~A.,  2006, \mn@doi [ApJ] {10.1086/505614}, \href
  {https://ui.adsabs.harvard.edu/abs/2006ApJ...647..910H} {647, 910}

\bibitem[\protect\citeauthoryear{{Hickson}}{{Hickson}}{1997}]{hickson1997}
{Hickson} P.,  1997, \mn@doi [ARAA] {10.1146/annurev.astro.35.1.357}, \href
  {https://ui.adsabs.harvard.edu/abs/1997ARA&A..35..357H} {35, 357}

\bibitem[\protect\citeauthoryear{{Ibata} et~al.,}{{Ibata}
  et~al.}{2017}]{ibata2017}
{Ibata} R.~A.,  et~al., 2017, \mn@doi [ApJ] {10.3847/1538-4357/aa855c}, \href
  {https://ui.adsabs.harvard.edu/abs/2017ApJ...848..128I} {848, 128}

\bibitem[\protect\citeauthoryear{{J{\'a}chym} et~al.,}{{J{\'a}chym}
  et~al.}{2019}]{jachym2019}
{J{\'a}chym} P.,  et~al., 2019, \mn@doi [ApJ] {10.3847/1538-4357/ab3e6c}, \href
  {https://ui.adsabs.harvard.edu/abs/2019ApJ...883..145J} {883, 145}

\bibitem[\protect\citeauthoryear{{Jaff{\'e}} et~al.,}{{Jaff{\'e}}
  et~al.}{2018}]{jaffe2018}
{Jaff{\'e}} Y.~L.,  et~al., 2018, \mn@doi [MNRAS] {10.1093/mnras/sty500}, \href
  {http://adsabs.harvard.edu/abs/2018MNRAS.tmp..484J} {}

\bibitem[\protect\citeauthoryear{{Kelly}}{{Kelly}}{2007}]{kelly2007}
{Kelly} B.~C.,  2007, \mn@doi [ApJ] {10.1086/519947}, \href
  {https://ui.adsabs.harvard.edu/abs/2007ApJ...665.1489K} {665, 1489}

\bibitem[\protect\citeauthoryear{{Kenney}, {van Gorkom}  \& {Vollmer}}{{Kenney}
  et~al.}{2004}]{kenney2004}
{Kenney} J.~D.~P.,  {van Gorkom} J.~H.,   {Vollmer} B.,  2004, \mn@doi [AJ]
  {10.1086/420805}, \href {http://adsabs.harvard.edu/abs/2004AJ....127.3361K}
  {127, 3361}

\bibitem[\protect\citeauthoryear{{Kenney}, {Abramson}  \&
  {Bravo-Alfaro}}{{Kenney} et~al.}{2015}]{kenney2015}
{Kenney} J.~D.~P.,  {Abramson} A.,   {Bravo-Alfaro} H.,  2015, \mn@doi [AJ]
  {10.1088/0004-6256/150/2/59}, \href
  {http://adsabs.harvard.edu/abs/2015AJ....150...59K} {150, 59}

\bibitem[\protect\citeauthoryear{{Larson}, {Tinsley}  \& {Caldwell}}{{Larson}
  et~al.}{1980}]{larson1980}
{Larson} R.~B.,  {Tinsley} B.~M.,   {Caldwell} C.~N.,  1980, \mn@doi [ApJ]
  {10.1086/157917}, \href {http://adsabs.harvard.edu/abs/1980ApJ...237..692L}
  {237, 692}

\bibitem[\protect\citeauthoryear{{Lee} \& {Chung}}{{Lee} \&
  {Chung}}{2018}]{lee2018}
{Lee} B.,  {Chung} A.,  2018, \mn@doi [ApJL] {10.3847/2041-8213/aae4d9}, \href
  {https://ui.adsabs.harvard.edu/abs/2018ApJ...866L..10L} {866, L10}

\bibitem[\protect\citeauthoryear{{Lee} et~al.,}{{Lee} et~al.}{2017}]{lee2017}
{Lee} B.,  et~al., 2017, \mn@doi [MNRAS] {10.1093/mnras/stw3162}, \href
  {https://ui.adsabs.harvard.edu/abs/2017MNRAS.466.1382L} {466, 1382}

\bibitem[\protect\citeauthoryear{{Lim}, {Mo}, {Lu}, {Wang}  \& {Yang}}{{Lim}
  et~al.}{2017}]{lim2017}
{Lim} S.~H.,  {Mo} H.~J.,  {Lu} Y.,  {Wang} H.,   {Yang} X.,  2017, \mn@doi
  [MNRAS] {10.1093/mnras/stx1462}, \href
  {http://adsabs.harvard.edu/abs/2017MNRAS.470.2982L} {470, 2982}

\bibitem[\protect\citeauthoryear{{Lotz}, {Primack}  \& {Madau}}{{Lotz}
  et~al.}{2004}]{lotz2004}
{Lotz} J.~M.,  {Primack} J.,   {Madau} P.,  2004, \mn@doi [AJ]
  {10.1086/421849}, \href
  {https://ui.adsabs.harvard.edu/abs/2004AJ....128..163L} {128, 163}

\bibitem[\protect\citeauthoryear{{McGee}, {Balogh}, {Bower}, {Font}  \&
  {McCarthy}}{{McGee} et~al.}{2009}]{mcgee2009}
{McGee} S.~L.,  {Balogh} M.~L.,  {Bower} R.~G.,  {Font} A.~S.,   {McCarthy}
  I.~G.,  2009, \mn@doi [MNRAS] {10.1111/j.1365-2966.2009.15507.x}, \href
  {http://adsabs.harvard.edu/abs/2009MNRAS.400..937M} {400, 937}

\bibitem[\protect\citeauthoryear{{McPartland}, {Ebeling}, {Roediger}  \&
  {Blumenthal}}{{McPartland} et~al.}{2016}]{mcpartland2016}
{McPartland} C.,  {Ebeling} H.,  {Roediger} E.,   {Blumenthal} K.,  2016,
  \mn@doi [MNRAS] {10.1093/mnras/stv2508}, \href
  {https://ui.adsabs.harvard.edu/abs/2016MNRAS.455.2994M} {455, 2994}

\bibitem[\protect\citeauthoryear{{Mihos} \& {Hernquist}}{{Mihos} \&
  {Hernquist}}{1994a}]{mihos1994a}
{Mihos} J.~C.,  {Hernquist} L.,  1994a, \mn@doi [ApJl] {10.1086/187299}, \href
  {http://adsabs.harvard.edu/abs/1994ApJ...425L..13M} {425, L13}

\bibitem[\protect\citeauthoryear{{Mihos} \& {Hernquist}}{{Mihos} \&
  {Hernquist}}{1994b}]{mihos1994b}
{Mihos} J.~C.,  {Hernquist} L.,  1994b, \mn@doi [ApJl] {10.1086/187460}, \href
  {http://adsabs.harvard.edu/abs/1994ApJ...431L...9M} {431, L9}

\bibitem[\protect\citeauthoryear{{Mohr}, {Mathiesen}  \& {Evrard}}{{Mohr}
  et~al.}{1999}]{mohr1999}
{Mohr} J.~J.,  {Mathiesen} B.,   {Evrard} A.~E.,  1999, \mn@doi [ApJ]
  {10.1086/307227}, \href
  {https://ui.adsabs.harvard.edu/abs/1999ApJ...517..627M} {517, 627}

\bibitem[\protect\citeauthoryear{{Moore}, {Katz}, {Lake}, {Dressler}  \&
  {Oemler}}{{Moore} et~al.}{1996}]{moore1996}
{Moore} B.,  {Katz} N.,  {Lake} G.,  {Dressler} A.,   {Oemler} A.,  1996,
  \mn@doi [Nature] {10.1038/379613a0}, \href
  {http://adsabs.harvard.edu/abs/1996Natur.379..613M} {379, 613}

\bibitem[\protect\citeauthoryear{{Moretti} et~al.,}{{Moretti}
  et~al.}{2020}]{moretti2020}
{Moretti} A.,  et~al., 2020, \mn@doi [ApJ] {10.3847/1538-4357/ab616a}, \href
  {https://ui.adsabs.harvard.edu/abs/2020ApJ...889....9M} {889, 9}

\bibitem[\protect\citeauthoryear{{Nulsen}}{{Nulsen}}{1982}]{nulsen1982}
{Nulsen} P.~E.~J.,  1982, \mn@doi [MNRAS] {10.1093/mnras/198.4.1007}, \href
  {https://ui.adsabs.harvard.edu/abs/1982MNRAS.198.1007N} {198, 1007}

\bibitem[\protect\citeauthoryear{{Oman}, {Bah{\'e}}, {Healy}, {Hess}, {Hudson}
  \& {Verheijen}}{{Oman} et~al.}{2021}]{oman2021}
{Oman} K.~A.,  {Bah{\'e}} Y.~M.,  {Healy} J.,  {Hess} K.~M.,  {Hudson} M.~J.,
  {Verheijen} M. A.~W.,  2021, \mn@doi [MNRAS] {10.1093/mnras/staa3845}, \href
  {https://ui.adsabs.harvard.edu/abs/2021MNRAS.501.5073O} {501, 5073}

\bibitem[\protect\citeauthoryear{{Pallero}, {G{\'o}mez}, {Padilla}, {Bah{\'e}},
  {Vega-Mart{\'\i}nez}  \& {Torres-Flores}}{{Pallero}
  et~al.}{2020}]{pallero2020}
{Pallero} D.,  {G{\'o}mez} F.~A.,  {Padilla} N.~D.,  {Bah{\'e}} Y.~M.,
  {Vega-Mart{\'\i}nez} C.~A.,   {Torres-Flores} S.,  2020, arXiv e-prints,
  \href {https://ui.adsabs.harvard.edu/abs/2020arXiv201208593P} {p.
  arXiv:2012.08593}

\bibitem[\protect\citeauthoryear{{Peng} et~al.,}{{Peng}
  et~al.}{2010}]{peng2010}
{Peng} Y.-j.,  et~al., 2010, \mn@doi [ApJ] {10.1088/0004-637X/721/1/193}, \href
  {http://adsabs.harvard.edu/abs/2010ApJ...721..193P} {721, 193}

\bibitem[\protect\citeauthoryear{{Peng}, {Maiolino}  \& {Cochrane}}{{Peng}
  et~al.}{2015}]{peng2015}
{Peng} Y.,  {Maiolino} R.,   {Cochrane} R.,  2015, \mn@doi [Nature]
  {10.1038/nature14439}, \href
  {http://adsabs.harvard.edu/abs/2015Natur.521..192P} {521, 192}

\bibitem[\protect\citeauthoryear{{Poggianti} et~al.,}{{Poggianti}
  et~al.}{2016}]{poggianti2016}
{Poggianti} B.~M.,  et~al., 2016, \mn@doi [AJ] {10.3847/0004-6256/151/3/78},
  \href {https://ui.adsabs.harvard.edu/abs/2016AJ....151...78P} {151, 78}

\bibitem[\protect\citeauthoryear{{Poggianti} et~al.,}{{Poggianti}
  et~al.}{2017}]{poggianti2017}
{Poggianti} B.~M.,  et~al., 2017, \mn@doi [ApJ] {10.3847/1538-4357/aa78ed},
  \href {http://adsabs.harvard.edu/abs/2017ApJ...844...48P} {844, 48}

\bibitem[\protect\citeauthoryear{{Poggianti} et~al.,}{{Poggianti}
  et~al.}{2019}]{poggianti2019_baryon}
{Poggianti} B.~M.,  et~al., 2019, \mn@doi [ApJ] {10.3847/1538-4357/ab5224},
  \href {https://ui.adsabs.harvard.edu/abs/2019ApJ...887..155P} {887, 155}

\bibitem[\protect\citeauthoryear{{Quilis}, {Moore}  \& {Bower}}{{Quilis}
  et~al.}{2000}]{quilis2000}
{Quilis} V.,  {Moore} B.,   {Bower} R.,  2000, \mn@doi [Science]
  {10.1126/science.288.5471.1617}, \href
  {http://adsabs.harvard.edu/abs/2000Sci...288.1617Q} {288, 1617}

\bibitem[\protect\citeauthoryear{{Ramos-Mart{\'\i}nez}, {G{\'o}mez}  \&
  {P{\'e}rez-Villegas}}{{Ramos-Mart{\'\i}nez}
  et~al.}{2018}]{ramos-martinez2018}
{Ramos-Mart{\'\i}nez} M.,  {G{\'o}mez} G.~C.,   {P{\'e}rez-Villegas} {\'A}.,
  2018, \mn@doi [MNRAS] {10.1093/mnras/sty393}, \href
  {https://ui.adsabs.harvard.edu/abs/2018MNRAS.476.3781R} {476, 3781}

\bibitem[\protect\citeauthoryear{{Rasmussen}, {Ponman}  \&
  {Mulchaey}}{{Rasmussen} et~al.}{2006}]{rasmussen2006}
{Rasmussen} J.,  {Ponman} T.~J.,   {Mulchaey} J.~S.,  2006, \mn@doi [MNRAS]
  {10.1111/j.1365-2966.2006.10492.x}, \href
  {https://ui.adsabs.harvard.edu/abs/2006MNRAS.370..453R} {370, 453}

\bibitem[\protect\citeauthoryear{{Rhee}, {Smith}, {Choi}, {Yi}, {Jaff{\'e}},
  {Candlish}  \& {S{\'a}nchez-J{\'a}nssen}}{{Rhee} et~al.}{2017}]{rhee2017}
{Rhee} J.,  {Smith} R.,  {Choi} H.,  {Yi} S.~K.,  {Jaff{\'e}} Y.,  {Candlish}
  G.,   {S{\'a}nchez-J{\'a}nssen} R.,  2017, \mn@doi [ApJ]
  {10.3847/1538-4357/aa6d6c}, \href
  {http://adsabs.harvard.edu/abs/2017ApJ...843..128R} {843, 128}

\bibitem[\protect\citeauthoryear{{Roberts} \& {Parker}}{{Roberts} \&
  {Parker}}{2017}]{roberts2017}
{Roberts} I.~D.,  {Parker} L.~C.,  2017, \mn@doi [MNRAS]
  {10.1093/mnras/stx317}, \href
  {http://adsabs.harvard.edu/abs/2017MNRAS.467.3268R} {467, 3268}

\bibitem[\protect\citeauthoryear{{Roberts} \& {Parker}}{{Roberts} \&
  {Parker}}{2020}]{roberts2020}
{Roberts} I.~D.,  {Parker} L.~C.,  2020, \mn@doi [MNRAS]
  {10.1093/mnras/staa1213}, \href
  {https://ui.adsabs.harvard.edu/abs/2020MNRAS.495..554R} {495, 554}

\bibitem[\protect\citeauthoryear{{Roberts}, {Parker}, {Brown}, {Joshi},
  {Hlavacek-Larrondo}  \& {Wadsley}}{{Roberts} et~al.}{2019}]{roberts2019}
{Roberts} I.~D.,  {Parker} L.~C.,  {Brown} T.,  {Joshi} G.~D.,
  {Hlavacek-Larrondo} J.,   {Wadsley} J.,  2019, \mn@doi [ApJ]
  {10.3847/1538-4357/ab04f7}, \href
  {https://ui.adsabs.harvard.edu/abs/2019ApJ...873...42R} {873, 42}

\bibitem[\protect\citeauthoryear{{Roberts} et~al.,}{{Roberts}
  et~al.}{2021a}]{roberts2021_LOFARclust}
{Roberts} I.~D.,  et~al., 2021a, \mn@doi [A\&A] {10.1051/0004-6361/202140784},
  \href {https://ui.adsabs.harvard.edu/abs/2021A&A...650A.111R} {650, A111}

\bibitem[\protect\citeauthoryear{{Roberts}, {van Weeren}, {McGee}, {Botteon},
  {Ignesti}  \& {Rottgering}}{{Roberts} et~al.}{2021b}]{roberts2021_LOFARgrp}
{Roberts} I.~D.,  {van Weeren} R.~J.,  {McGee} S.~L.,  {Botteon} A.,  {Ignesti}
  A.,   {Rottgering} H.~J.~A.,  2021b, A\&A, 652, A153

\bibitem[\protect\citeauthoryear{{Rodriguez Wimberly}, {Cooper}, {Fillingham},
  {Boylan-Kolchin}, {Bullock}  \& {Garrison-Kimmel}}{{Rodriguez Wimberly}
  et~al.}{2019}]{rodriguez-wimberly2019}
{Rodriguez Wimberly} M.~K.,  {Cooper} M.~C.,  {Fillingham} S.~P.,
  {Boylan-Kolchin} M.,  {Bullock} J.~S.,   {Garrison-Kimmel} S.,  2019, \mn@doi
  [MNRAS] {10.1093/mnras/sty3357}, \href
  {https://ui.adsabs.harvard.edu/abs/2019MNRAS.483.4031R} {483, 4031}

\bibitem[\protect\citeauthoryear{{Salim} et~al.,}{{Salim}
  et~al.}{2016}]{salim2016}
{Salim} S.,  et~al., 2016, \mn@doi [ApJS] {10.3847/0067-0049/227/1/2}, \href
  {http://adsabs.harvard.edu/abs/2016ApJS..227....2S} {227, 2}

\bibitem[\protect\citeauthoryear{{Salim}, {Boquien}  \& {Lee}}{{Salim}
  et~al.}{2018}]{salim2018}
{Salim} S.,  {Boquien} M.,   {Lee} J.~C.,  2018, \mn@doi [ApJ]
  {10.3847/1538-4357/aabf3c}, \href
  {https://ui.adsabs.harvard.edu/abs/2018ApJ...859...11S} {859, 11}

\bibitem[\protect\citeauthoryear{Scholz \& Stephens}{Scholz \&
  Stephens}{1987}]{scholz1987}
Scholz F.~W.,  Stephens M.~A.,  1987, \mn@doi [Journal of the American
  Statistical Association] {10.1080/01621459.1987.10478517}, 82, 918

\bibitem[\protect\citeauthoryear{{Shimwell} et~al.,}{{Shimwell}
  et~al.}{2017}]{shimwell2017}
{Shimwell} T.~W.,  et~al., 2017, \mn@doi [A\&A] {10.1051/0004-6361/201629313},
  \href {https://ui.adsabs.harvard.edu/abs/2017A&A...598A.104S} {598, A104}

\bibitem[\protect\citeauthoryear{{Shimwell} et~al.,}{{Shimwell}
  et~al.}{2019}]{shimwell2019}
{Shimwell} T.~W.,  et~al., 2019, \mn@doi [A\&A] {10.1051/0004-6361/201833559},
  \href {https://ui.adsabs.harvard.edu/abs/2019A&A...622A...1S} {622, A1}

\bibitem[\protect\citeauthoryear{{Smith} et~al.,}{{Smith}
  et~al.}{2010}]{smith2010}
{Smith} G.~P.,  et~al., 2010, \mn@doi [MNRAS]
  {10.1111/j.1365-2966.2010.17311.x}, \href
  {http://adsabs.harvard.edu/abs/2010MNRAS.409..169S} {409, 169}

\bibitem[\protect\citeauthoryear{{Steinhauser}, {Haider}, {Kapferer}  \&
  {Schindler}}{{Steinhauser} et~al.}{2012}]{steinhauser2012}
{Steinhauser} D.,  {Haider} M.,  {Kapferer} W.,   {Schindler} S.,  2012,
  \mn@doi [A\&A] {10.1051/0004-6361/201118311}, \href
  {https://ui.adsabs.harvard.edu/abs/2012A&A...544A..54S} {544, A54}

\bibitem[\protect\citeauthoryear{{Sun}, {Donahue}, {Roediger}, {Nulsen},
  {Voit}, {Sarazin}, {Forman}  \& {Jones}}{{Sun} et~al.}{2010}]{sun2010}
{Sun} M.,  {Donahue} M.,  {Roediger} E.,  {Nulsen} P.~E.~J.,  {Voit} G.~M.,
  {Sarazin} C.,  {Forman} W.,   {Jones} C.,  2010, \mn@doi [ApJ]
  {10.1088/0004-637X/708/2/946}, \href
  {https://ui.adsabs.harvard.edu/abs/2010ApJ...708..946S} {708, 946}

\bibitem[\protect\citeauthoryear{{Sun} et~al.,}{{Sun} et~al.}{2021}]{sun2021}
{Sun} M.,  et~al., 2021, arXiv e-prints, \href
  {https://ui.adsabs.harvard.edu/abs/2021arXiv210309205S} {p. arXiv:2103.09205}

\bibitem[\protect\citeauthoryear{{Tomi{\v{c}}i{\'c}}
  et~al.,}{{Tomi{\v{c}}i{\'c}} et~al.}{2018}]{tomicic2018}
{Tomi{\v{c}}i{\'c}} N.,  et~al., 2018, \mn@doi [ApJL]
  {10.3847/2041-8213/aaf810}, \href
  {https://ui.adsabs.harvard.edu/abs/2018ApJ...869L..38T} {869, L38}

\bibitem[\protect\citeauthoryear{{Troncoso Iribarren}, {Padilla}, {Contreras},
  {Rodriguez}, {Garc{\'i}a-Lambas}  \& {Lagos}}{{Troncoso Iribarren}
  et~al.}{2016}]{troncoso-iribarren2016}
{Troncoso Iribarren} P.,  {Padilla} N.,  {Contreras} S.,  {Rodriguez} S.,
  {Garc{\'i}a-Lambas} D.,   {Lagos} C.,  2016, \mn@doi [Galaxies]
  {10.3390/galaxies4040077}, \href
  {https://ui.adsabs.harvard.edu/abs/2016Galax...4...77T} {4, 77}

\bibitem[\protect\citeauthoryear{{Troncoso-Iribarren}, {Padilla}, {Santander},
  {Lagos}, {Garc{\'\i}a-Lambas}, {Rodr{\'\i}guez}  \&
  {Contreras}}{{Troncoso-Iribarren} et~al.}{2020}]{troncoso-iribarren2020}
{Troncoso-Iribarren} P.,  {Padilla} N.,  {Santander} C.,  {Lagos} C.~D.~P.,
  {Garc{\'\i}a-Lambas} D.,  {Rodr{\'\i}guez} S.,   {Contreras} S.,  2020,
  \mn@doi [MNRAS] {10.1093/mnras/staa274}, \href
  {https://ui.adsabs.harvard.edu/abs/2020MNRAS.497.4145T} {497, 4145}

\bibitem[\protect\citeauthoryear{{Vollmer}, {Soida}, {Chung}, {Chemin},
  {Braine}, {Boselli}  \& {Beck}}{{Vollmer} et~al.}{2009}]{vollmer2009}
{Vollmer} B.,  {Soida} M.,  {Chung} A.,  {Chemin} L.,  {Braine} J.,  {Boselli}
  A.,   {Beck} R.,  2009, \mn@doi [A\&A] {10.1051/0004-6361/200811140}, \href
  {https://ui.adsabs.harvard.edu/abs/2009A&A...496..669V} {496, 669}

\bibitem[\protect\citeauthoryear{{Vollmer} et~al.,}{{Vollmer}
  et~al.}{2012}]{vollmer2012}
{Vollmer} B.,  et~al., 2012, \mn@doi [A\&A] {10.1051/0004-6361/201117680},
  \href {https://ui.adsabs.harvard.edu/abs/2012A&A...537A.143V} {537, A143}

\bibitem[\protect\citeauthoryear{{Vulcani} et~al.,}{{Vulcani}
  et~al.}{2018a}]{vulcani2018a}
{Vulcani} B.,  et~al., 2018a, \mn@doi [MNRAS] {10.1093/mnras/sty2095}, \href
  {https://ui.adsabs.harvard.edu/abs/2018MNRAS.480.3152V} {480, 3152}

\bibitem[\protect\citeauthoryear{{Vulcani} et~al.,}{{Vulcani}
  et~al.}{2018b}]{vulcani2018b}
{Vulcani} B.,  et~al., 2018b, \mn@doi [ApJL] {10.3847/2041-8213/aae68b}, \href
  {https://ui.adsabs.harvard.edu/abs/2018ApJ...866L..25V} {866, L25}

\bibitem[\protect\citeauthoryear{{Wang}, {Xu}, {Lee}, {Du}, {Overzier}  \&
  {Shao}}{{Wang} et~al.}{2020}]{wang2020}
{Wang} J.,  {Xu} W.,  {Lee} B.,  {Du} M.,  {Overzier} R.,   {Shao} L.,  2020,
  arXiv e-prints, \href {https://ui.adsabs.harvard.edu/abs/2020arXiv200908159W}
  {p. arXiv:2009.08159}

\bibitem[\protect\citeauthoryear{{Wetzel}, {Tinker}  \& {Conroy}}{{Wetzel}
  et~al.}{2012}]{wetzel2012}
{Wetzel} A.~R.,  {Tinker} J.~L.,   {Conroy} C.,  2012, \mn@doi [MNRAS]
  {10.1111/j.1365-2966.2012.21188.x}, \href
  {http://adsabs.harvard.edu/abs/2012MNRAS.424..232W} {424, 232}

\bibitem[\protect\citeauthoryear{{Wetzel}, {Tollerud}  \& {Weisz}}{{Wetzel}
  et~al.}{2015}]{wetzel2015}
{Wetzel} A.~R.,  {Tollerud} E.~J.,   {Weisz} D.~R.,  2015, \mn@doi [ApJL]
  {10.1088/2041-8205/808/1/L27}, \href
  {http://adsabs.harvard.edu/abs/2015ApJ...808L..27W} {808, L27}

\bibitem[\protect\citeauthoryear{{Yang}, {Mo}, {van den Bosch}  \&
  {Jing}}{{Yang} et~al.}{2005}]{yang2005}
{Yang} X.,  {Mo} H.~J.,  {van den Bosch} F.~C.,   {Jing} Y.~P.,  2005, \mn@doi
  [MNRAS] {10.1111/j.1365-2966.2005.08560.x}, \href
  {http://adsabs.harvard.edu/abs/2005MNRAS.356.1293Y} {356, 1293}

\bibitem[\protect\citeauthoryear{{Yang}, {Mo}, {van den Bosch}, {Pasquali},
  {Li}  \& {Barden}}{{Yang} et~al.}{2007}]{yang2007}
{Yang} X.,  {Mo} H.~J.,  {van den Bosch} F.~C.,  {Pasquali} A.,  {Li} C.,
  {Barden} M.,  2007, \mn@doi [ApJ] {10.1086/522027}, \href
  {http://adsabs.harvard.edu/abs/2007ApJ...671..153Y} {671, 153}

\bibitem[\protect\citeauthoryear{{Yoon}, {Chung}, {Smith}  \&
  {Jaff{\'e}}}{{Yoon} et~al.}{2017}]{yoon2017}
{Yoon} H.,  {Chung} A.,  {Smith} R.,   {Jaff{\'e}} Y.~L.,  2017, \mn@doi [ApJ]
  {10.3847/1538-4357/aa6579}, \href
  {https://ui.adsabs.harvard.edu/abs/2017ApJ...838...81Y} {838, 81}

\bibitem[\protect\citeauthoryear{{Yun} et~al.,}{{Yun} et~al.}{2019}]{yun2019}
{Yun} K.,  et~al., 2019, \mn@doi [MNRAS] {10.1093/mnras/sty3156}, \href
  {https://ui.adsabs.harvard.edu/\#abs/2019MNRAS.483.1042Y} {483, 1042}

\bibitem[\protect\citeauthoryear{{Zabel} et~al.,}{{Zabel}
  et~al.}{2019}]{zabel2019}
{Zabel} N.,  et~al., 2019, \mn@doi [\mnras] {10.1093/mnras/sty3234}, \href
  {https://ui.adsabs.harvard.edu/abs/2019MNRAS.483.2251Z} {483, 2251}

\bibitem[\protect\citeauthoryear{{von der Linden}, {Wild}, {Kauffmann}, {White}
   \& {Weinmann}}{{von der Linden} et~al.}{2010}]{vonderlinden2010}
{von der Linden} A.,  {Wild} V.,  {Kauffmann} G.,  {White} S.~D.~M.,
  {Weinmann} S.,  2010, \mn@doi [MNRAS] {10.1111/j.1365-2966.2010.16375.x},
  \href {http://adsabs.harvard.edu/abs/2010MNRAS.404.1231V} {404, 1231}

\makeatother
\end{thebibliography}



\appendix
\onecolumn

\section{Ram Pressure Candidate Images} \label{sec:appendix_img}

\begin{longtable}{c c c c l}
\caption{Ram Pressure Candidates} \\
\toprule
Galaxy & RA & Dec & $z$ & Classification Notes \\
Number & [deg] & [deg] & & \\
\midrule
\endfirsthead
\toprule
Galaxy & RA & Dec & $z$ & Classification Notes \\
Number & [deg] & [deg] & & \\
\midrule
\endhead
\midrule
\multicolumn{4}{r}{{Continued on next page}} \\
\bottomrule
\endfoot
\bottomrule
\endlastfoot
1 & 123.6493 & 58.3661 & 0.0293 & Asymmetric $u$ to the west \\
2 & 124.5442 & 57.7588 & 0.0268 & Tails to the southeast, bright $u$ to the northwest \\
3 & 125.0641 & 56.3767 & 0.0300 & Tails to the south \\
4 & 125.1572 & 58.0453 & 0.0265 & Tail to the south \\
5 & 128.3571 & 41.0544 & 0.0250 & Tail to the west \\
6 & 130.0765 & 40.7965 & 0.0307 & Tails and bright $u$ to the south \\
7 & 136.8169 & 37.2154 & 0.0238 & Asymmetric $u$ to the north \\
8 & 137.3513 & 37.6858 & 0.0236 & Tails to the southeast, $u$ sources to the northwest \\
9 & 138.5934 & 30.4330 & 0.0215 & Tail to the east, bright $u$ on western edge \\
10 & 139.8783 & 38.9660 & 0.0278 & Tails to the southeast, bow feature on northwest edge? \\
11 & 140.0890 & 34.2384 & 0.0246 & Tail to the north, bright $u$ source on southern edge \\
12 & 143.6829 & 33.9727 & 0.0289 & Tail to the northeast \\
13 & 144.1025 & 31.8513 & 0.0222 & Asymmetric $u$ to northeast, tail to the south?\\
14 & 147.3384 & 34.6568 & 0.0400 & Tail to the north \\
15 & 147.7212 & 34.6549 & 0.0399 & Tail to the north \\
16 & 154.7025 & 38.4697 & 0.0223 & Tail to the south, asymmetric $u$ on northern edge (tidal?) \\
17 & 160.0280 & 39.1089 & 0.0342 & Tails to the southwest, asymmetric $u$ in the north \\
18 & 160.0464 & 38.5785 & 0.0360 & Tail to the southwest \\
19 & 160.2714 & 37.3860 & 0.0234 & Tail to the north \\
20 & 160.7922 & 39.0387 & 0.0339 & Asymmetric $u$, bow feature on western edge \\
21 & 161.0554 & 39.2021 & 0.0356 & Asymmetric $u$ to the east \\
22 & 161.4409 & 39.0194 & 0.0359 & Tail to the south, faint tail to the north? (tidal?) \\
23 & 161.6613 & 38.3612 & 0.0351 & Tail to the south\\
24 & 161.8615 & 38.9364 & 0.0350 & Tail to the east, bright $u$ on western edge \\
25 & 165.2576 & 46.5653 & 0.0222 & Asymmetric $u$ to the northeast \\
26 & 165.2820 & 45.6527 & 0.0248 & Tail to the southwest \\
27 & 168.7345 & 30.8958 & 0.0274 & Asymmetric $u$ to the east \\
28 & 172.0338 & 35.3783 & 0.0377 & Asymmetric $u$ to the southeast \\
29 & 173.0356 & 35.5109 & 0.0385 & Asymmetric $u$ to the west \\
30 & 173.0633 & 35.5125 & 0.0377 & Tails to the south \\
31 & 175.7723 & 33.8349 & 0.0315 & Tail to the south \\
32 & 175.8477 & 32.8801 & 0.0343 & Bow feature on eastern edge, faint tail to the northwest? \\
33 & 175.9726 & 33.3490 & 0.0322 & Tail to the northeast, bright $u$ on southwestern edge \\
34 & 176.4507 & 33.1483 & 0.0321 & Tail to the south, bright $u$ on northern edge \\
35 & 176.5412 & 33.2083 & 0.0329 & Tail to the west \\
36 & 197.5338 & 34.9055 & 0.0356 & Asymmetric $u$ to the south \\
37 & 197.6735 & 34.8172 & 0.0367 & Tail/asymmetric $u$ to the south \\
38 & 200.0656 & 33.6637 & 0.0391 & Tail to the north \\
39 & 200.7128 & 31.8259 & 0.0178 & Asymmetric $u$ to the northwest \\
40 & 200.8339 & 32.0635 & 0.0167 & Tails to the southeast \\
41 & 201.2553 & 36.4372 & 0.0195 & Asymmetric $u$ to the north \\
42 & 201.4014 & 36.3812 & 0.0188 & Tails to the north/northeast (tidal?) \\
43 & 231.9340 & 43.0689 & 0.0175 & Tail to the southwest \\
44 & 244.8756 & 37.7874 & 0.0327 & Asymmetric $u$ to the west \\
45 & 244.9793 & 38.1306 & 0.0304 & Asymmetric $u$ to the west \\
46 & 245.5684 & 38.4499 & 0.0292 & Asymmetric $u$ to the southeast \\
47 & 245.8868 & 38.0366 & 0.0338 & Tail to the southeast \\
48 & 248.7128 & 36.7517 & 0.0337 & Tail to the east \\
\label{tab:jellyfish_sample}
\end{longtable}

\begin{figure*}
    \centering
    \includegraphics[width=\textwidth]{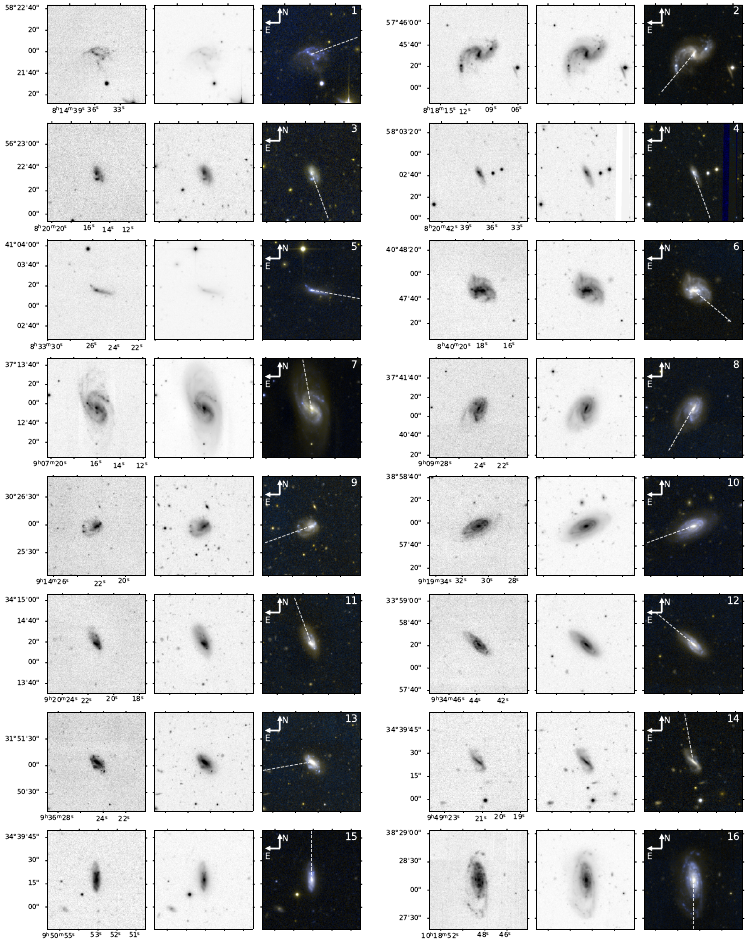}
    \caption{$u$-band (left), $r$-band (centre), and rgb (right) for the ram pressure candidates identified in this work.  White line indicates the estimated `tail' direction for each ram pressure candidate (see Section~\ref{sec:orient}).}
    \label{fig:RPimages1}
\end{figure*}

\begin{figure*}
    \centering
    \includegraphics[width=\textwidth]{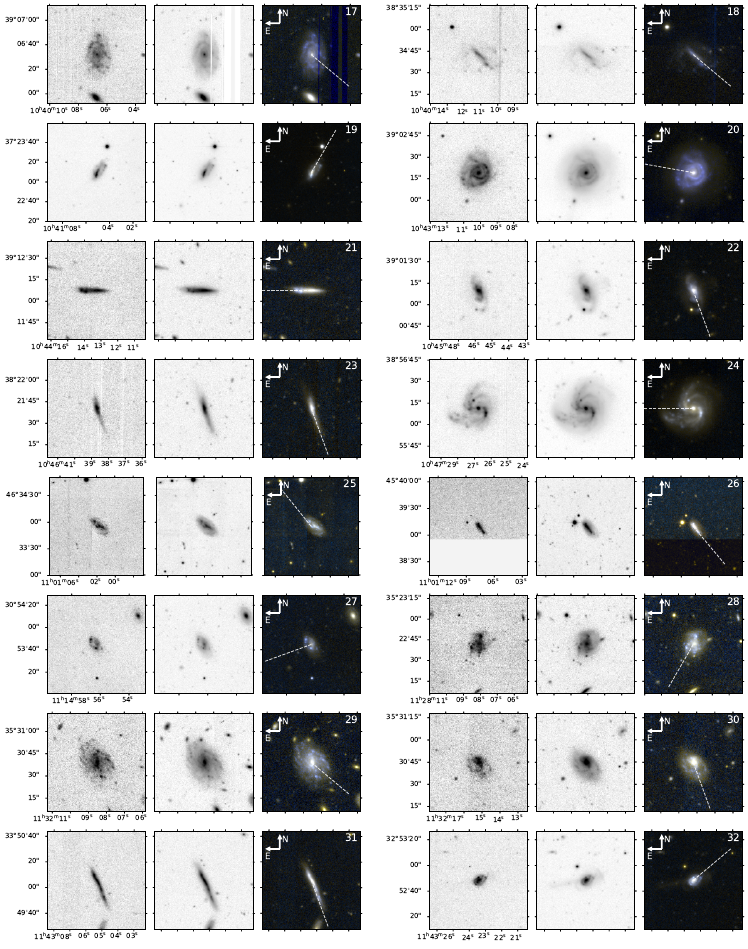}
    \caption{$u$-band (left), $r$-band (centre), and rgb (right) for the ram pressure candidates identified in this work.  White line indicates the estimated `tail' direction for each ram pressure candidate (see Section~\ref{sec:orient}).}
    \label{fig:RPimages2}
\end{figure*}

\begin{figure*}
    \centering
    \includegraphics[width=\textwidth]{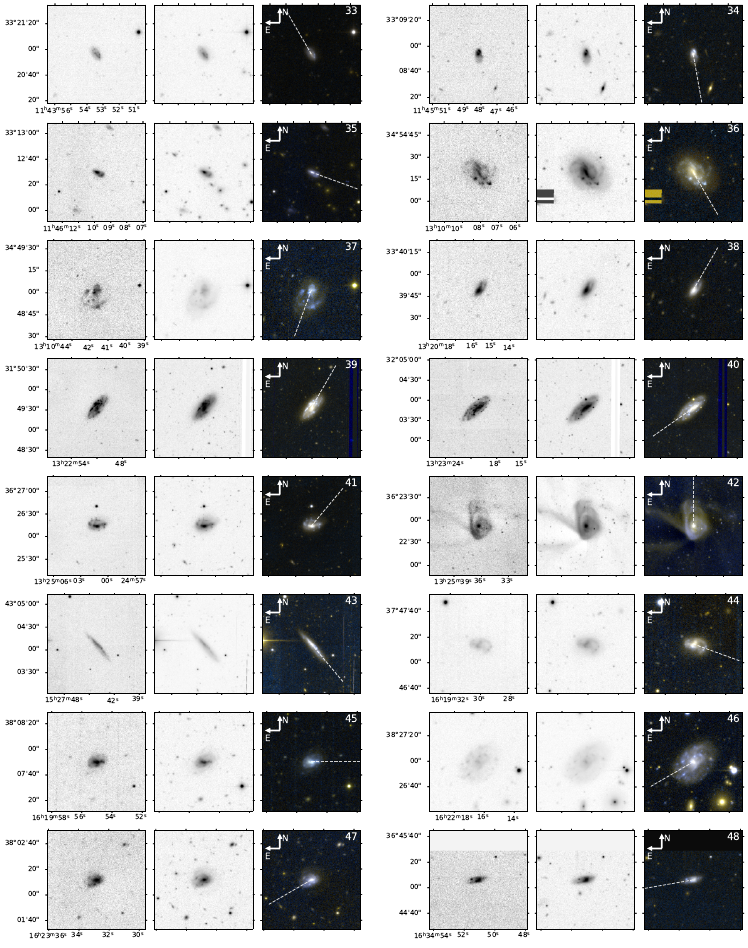}
    \caption{$u$-band (left), $r$-band (centre), and rgb (right) for the ram pressure candidates identified in this work.  White line indicates the estimated `tail' direction for each ram pressure candidate (see Section~\ref{sec:orient}).}
    \label{fig:RPimages3}
\end{figure*}


\bsp	
\label{lastpage}
\end{document}